\documentclass[acmsmall,screen,authorversion,nonacm]{acmart}

\usepackage{amsmath}
\usepackage{float}
\usepackage[table]{xcolor}
\usepackage{tabularx}
\usepackage{enumitem}
\usepackage{tikz}
\usepgflibrary{shapes}

\definecolor{EBCBelow}{HTML}{D55E00}
\definecolor{EBCNeutral}{HTML}{4D4D4D}
\definecolor{EBCAbove}{HTML}{0072B2}

\newcommand{\ebcVeryLow}{%
  \textcolor{EBCBelow}{\textbf{\texttt{- -}}}}
\newcommand{\ebcLow}{%
  \textcolor{EBCBelow}{\textbf{\texttt{-}}}}
\newcommand{\ebcTypical}{%
  \textcolor{EBCNeutral}{\textbf{\texttt{=}}}}
\newcommand{\ebcHigh}{%
  \textcolor{EBCAbove}{\textbf{\texttt{+}}}}
\newcommand{\ebcVeryHigh}{%
  \textcolor{EBCAbove}{\textbf{\texttt{+ +}}}}

\AtBeginDocument{%
  }

\setcopyright{acmlicensed}
\copyrightyear{2026}
\acmYear{2026}
\acmDOI{XXXXXXX.XXXXXXX}
\acmISBN{978-1-4503-XXXX-X/2018/06}

\begin{document}

\title{The EBC Proportionality Law: A Recurring Architectural-Role Pattern Across Java Software}

\author{Dzikri Rahadian Fudholi}
\email{dzikri.r.f@ugm.ac.id}
\orcid{0000-0001-9759-2353}
\author{Andrea Capiluppi}
\email{a.capiluppi@rug.nl}
\orcid{0000-0001-9469-6050}
\affiliation{%
  \institution{University of Groningen}
  \city{Groningen}
  \state{Groningen}
  \country{The Netherlands}
}


\begin{abstract}
Architectural roles describe how responsibilities are distributed within software systems. 
Entity--Boundary--Control (EBC) separates domain information, external interaction, and workflow coordination, but its proportions across diverse software remain unclear. 
This study examines file-level classifier-derived EBC proportionality in 1,556,985 Java files from 732 project instances and 78 functional categories.
A classifier trained on labels generated by a large language model assigned one representative role per file. 
The labels were aggregated at project and functional-category levels and analysed using descriptive statistics and density-based clustering.
The results show that Entity was less common than Boundary and Control.
The dominant cluster covered 48 categories, 472 project instances, and 47.75\% of classified files; its mean composition was 13.25\% Entity, 42.65\% Boundary, and 44.10\% Control.
The remaining 30 categories form one alternative and 11 smaller specialised patterns.
Some selected variations were consistent with category responsibilities, while others received mixed support. 
Agreement with independent expert labels was limited (31.37\% accuracy; macro-F1 0.31; Entity recall 13\%). 
All reported proportions and clusters are therefore patterns in \emph{classifier-assigned} roles, not verified architectural ground truth.
Within this measurement boundary, the study provides a large-scale empirical reference of EBC distributions at the file level, and it documents the limitations of the automated approach.
\end{abstract}




\maketitle

\section{Introduction}
\label{S1:Intro}

Software architecture assigns responsibilities to the elements of a software system.
These responsibilities determine which elements maintain information, communicate with external actors, and coordinate system behaviour.
A clear assignment of responsibilities helps connect system requirements with an understandable implementation structure~\cite{hofmeister2000applied,bass2021software}.
The distribution of these responsibilities therefore provides one view of how a software system is organised.

The Entity--Boundary--Control (EBC) model provides an established way to separate software responsibilities.
\citet{jacobson1993object} introduced EBC within use-case-driven object-oriented software engineering.
An Entity represents domain information and maintains system state, a Boundary manages communication with external actors, and a Control coordinates the actions required to realise system behaviour~\cite{jacobson1993object,jacobson1999unified}.
Together, these roles connect functional requirements with the software elements responsible for information, interaction, and coordination.

EBC can be examined through individual labels and their overall distribution.
Previous studies classified individual methods or classes and then counted how often each role appeared~\cite{dragan2010automatic,hoquang2022role}.
In this study, role assignment gives each Java file one Entity, Boundary, or Control label.
Classifier-derived EBC proportionality then describes the percentage of files assigned to each role.
The three percentages form compositional data because they represent parts of the same whole and sum to 100\%~\cite{aitchison1986statistical}.
Project-level proportionality describes one project instance, while category-level proportionality describes projects with a similar functional purpose.
This representation allows the EBC composition of different projects and functional categories to be compared.

Classifier-derived EBC proportionality may contain both commonality and functional variation.
Software projects share general architectural responsibilities, but their functional purposes may place different emphasis on information, external interaction, and workflow coordination.
Software classifications provide context for these differences by separating projects according to the primary functions they perform~\cite{forward2008taxonomy}.
Project diversity is also necessary when empirical findings are expected to apply beyond a small collection of software systems~\cite{nagappan2013diversity}.
Functional categories therefore provide a basis for examining whether Classifier-derived EBC proportionality remains similar across different kinds of Java software.

Existing empirical research has mainly examined the identification and evolution of architectural roles in source code.
Previous approaches identified role stereotypes at method and class levels using rule-based and automated classification techniques~\cite{dragan2010automatic,moreno2012jstereocode}.
More recent research used machine learning to classify 1,547 Java classes from three projects and examined how their role distributions changed across software versions~\cite{hoquang2022role}.
These studies establish that architectural roles can be identified and counted, but they do not examine whether file-level Classifier-derived EBC proportions recur across a large and functionally diverse collection of Java projects.
Evidence therefore remains limited on the overall distribution, recurrence, and functional variation of Classifier-derived EBC proportionality.

This study addresses this gap through a file-level analysis of Java software.
The dataset was curated from the Awesome Java collection~\cite{awesomejava} and contains 732 project instances across 78 functional categories.
The analytical corpus contains 1,556,985 Java source files that received an Entity, Boundary, or Control label.
The annotation procedure combines large-language-model-generated weak labels with a LightGBM classifier to assign one representative EBC role to each Java source file.
The classifier predictions are compared with independently assigned expert labels to document the limitations of the automated measurement.
The resulting file labels are then aggregated and analysed at project and functional-category levels.

The empirical analysis is organised through three research questions.

\begin{enumerate}

\item \textbf{RQ1: How is Classifier-derived EBC proportionality distributed across projects and software functional categories?}

\textit{Rationale.}
RQ1 establishes the central values and variation of Entity, Boundary, and Control proportions.
These distributions provide the empirical basis for examining recurrence and variation in the following questions.

\item \textbf{RQ2: To what extent does Classifier-derived EBC proportionality exhibit a recurring pattern across Java software?}

\textit{Rationale.}
RQ2 determines whether a substantial group of functional categories shares a similar EBC composition.
It also measures the coverage of this composition across functional categories, project instances, and Java source files.

\item \textbf{RQ3: What variations from the recurring Classifier-derived EBC proportionality occur across software functional categories, and are these variations consistent with their architectural characteristics?}

\textit{Rationale.}
RQ3 identifies the patterns outside the recurring EBC composition.
It also examines whether selected variations are consistent with the functions and architectural characteristics of their software categories.

\end{enumerate}

The results identify a recurring but non-universal EBC composition.
The dominant cluster contains 48 of the 78 functional categories and 472 of the 732 project instances, representing 47.75\% of the classified Java files.
Its mean composition is 13.25\% Entity, 42.65\% Boundary, and 44.10\% Control.
This recurring composition is defined as the \emph{EBC Proportionality Law}.
The remaining categories form one alternative pattern and several smaller specialised patterns, showing that functional variation remains present.
The Law therefore describes an empirical reference within the studied Java corpus rather than a universal or mandatory architectural rule.

This study makes three main contributions.
First, it provides a large-scale empirical characterisation of file-level Classifier-derived EBC proportions across 732 Java project instances and 78 functional categories.
Second, it introduces the Classifier-derived EBC proportionality Law and quantifies its coverage across functional categories, project instances, and Java source files.
Third, it identifies alternative and specialised EBC patterns and examines selected variations in relation to their architectural characteristics.
The accompanying replication package documents the data, automated annotation procedure, and analytical process used to produce these findings.

The remainder of the paper develops and evaluates this empirical finding.
Section~2 presents the research background and related work.
Section~3 describes the study design, dataset, automated annotation procedure, and analysis methods.
Section~4 presents the empirical results.
Section~5 discusses the findings and their implications for researchers, practitioners, and software engineering educators.
Section~6 presents the threats to validity, and Section~7 concludes the paper.

\section{Background and Related Work}
\label{S2:Background}
Understanding EBC proportionality requires a common foundation for architectural roles, software functionality, automated classification, and recurring empirical patterns.
This section first introduces Entity--Boundary--Control and related approaches for separating software responsibilities.
It then explains how functional categories provide context for comparing different kinds of software.
Next, it reviews automated methods for identifying architectural roles at different source-code levels.
Finally, it discusses recurring empirical patterns in software engineering and establishes the research gap addressed by this study.

\subsection{Entity–Boundary–Control and Architectural Roles}
\label{S2:EBC}
Architectural responsibilities describe the work assigned to different elements of a software system. 
Software architecture organises these elements and defines how they interact to satisfy system requirements~\cite{hofmeister2000applied, medvidovic2010software}. 
Responsibility-driven design further describes software objects according to the information they maintain, the behaviour they perform, and the interactions they support~\cite{wirfsbrock2003object, larman2012applying}. 
A clear assignment of responsibilities separates different forms of software work and makes the system structure easier to understand. 
Architectural responsibilities therefore provide the basis for examining how Entity, Boundary, and Control roles are distributed across a software system.

The Entity–Boundary–Control paradigm separates software responsibilities into three architectural roles. 
Jacobson introduced the paradigm within use-case-driven object-oriented software engineering to identify the objects involved in realising a use case~\cite{jacobson1993object}. 
The approach connects required system behaviour with analysis objects by assigning each object an Entity, Boundary, or Control role~\cite{jacobson1993object}. 
The Unified Software Development Process later continued this role-based approach within iterative software development~\cite{jacobson1999unified}. 
EBC therefore provides a structured link between functional requirements and the responsibilities assigned to software elements.

Each EBC role represents a different type of software responsibility. 
An Entity represents domain information and maintains state that remains relevant to the system~\cite{jacobson1993object,jacobson1999unified}. 
A Boundary manages communication between the system and external actors, including users, devices, and other software systems~\cite{jacobson1993object,jacobson1999unified}. 
A Control coordinates the actions required to realise a use case and connects Boundary objects with Entity objects~\cite{jacobson1993object,jacobson1999unified}. 
The three roles separate information, external interaction, and behavioural coordination while allowing them to collaborate in the same system behaviour.

EBC is related to other approaches that separate software responsibilities. 
The Model–View–Controller pattern divides an interactive application into Model, View, and Controller elements~\cite{reenskaug1979models,krasner1988cookbook}. 
The Model represents application information and domain state, the View presents information to the user, and the Controller manages user interaction with the Model and View~\cite{krasner1988cookbook}. 
This division separates domain representation, presentation, and user interaction. 
MVC therefore provides a useful reference point for understanding the separation of responsibilities in EBC.

The correspondence between MVC and EBC is approximate rather than direct. 
The MVC Model and EBC Entity both represent domain information, although the Model may also contain application behaviour~\cite{reenskaug1979models,jacobson1993object}. 
The MVC View focuses on visual presentation, whereas an EBC Boundary covers communication with users, devices, and external software systems~\cite{krasner1988cookbook,jacobson1993object}. 
The MVC Controller manages user interaction, whereas an EBC Control coordinates the broader sequence of actions required to realise a use case~\cite{krasner1988cookbook,jacobson1993object}. 
EBC and MVC therefore share a separation-of-responsibilities principle but apply that principle to different architectural concerns.

Architectural roles can be examined at different levels of source-code organisation. 
EBC originally assigns Entity, Boundary, and Control responsibilities to analysis objects and classes~\cite{jacobson1993object,jacobson1999unified}. 
Later studies have identified role stereotypes at method and class levels to recover design information from source code~\cite{dragan2010automatic,moreno2012jstereocode,hoquang2022role}. 
A class may contain more than one responsibility, while a Java source file may contain several classes and methods with different responsibilities~\cite{hoquang2022role}. 
The selected level of analysis can therefore influence how architectural roles are identified and counted. 
This granularity issue represents an important limitation for studies that measure EBC proportionality at the source-file level.

Existing empirical research has mainly examined the identification and evolution of architectural role stereotypes.
\citet{dragan2010automatic} inferred class stereotypes from the frequency and distribution of method stereotypes, using Entity, Boundary, and Control roles as the basis of the classification.
\citet{moreno2012jstereocode} introduced JStereoCode to identify method and class stereotypes automatically in Java source code.
\citet{hoquang2022role} used machine learning to classify 1,547 Java classes from three projects and examined how role-stereotype ratios changed across software versions.
Together, these studies show that role distributions can support software design analysis, but their scope remains limited in terms of projects and functional categories.
Evidence therefore remains limited on whether Classifier-derived EBC proportionality exhibits recurring patterns across a large and functionally diverse collection of Java projects.

\subsection{Software Functional Categories}
\label{S2:FunctionalCategories}
Software systems are developed for different functional purposes, and these purposes create different architectural needs.
\citet{forward2008taxonomy} organised software into types such as data-dominant, system, control-dominant, and computation-dominant software to make the context of empirical evidence more explicit.
Different types of software emphasise different requirements, components, and interactions, which can influence their architectural structures~\cite{bass2021software}.
Functional categories therefore provide a useful basis for examining whether the balance among Entity, Boundary, and Control roles remains similar across different kinds of Java software.

Functional categories can be assigned through automated analysis or manual curation.
\citet{kawaguchi2006mudablue} introduced MUDABlue, which used latent semantic analysis of source code to categorise open-source projects automatically.
Curated collections instead organise projects through maintainer review; Awesome Java, for example, groups Java frameworks, libraries, and software under named functional categories~\cite{awesomejava}.
Both approaches provide functional context for comparing projects that serve similar purposes, although a category describes the primary purpose of a project rather than its internal architecture.

Functional categories are important when empirical findings are expected to apply beyond individual projects.
\citet{nagappan2013diversity} showed that a large project sample does not by itself ensure generality when the selected projects cover only a narrow range of software types.
Comparing classifier-derived EBC proportions across functional categories makes it possible to determine whether a recurring pattern is widespread or concentrated in particular kinds of software.
It also provides the context needed to interpret category-level variations according to the architectural responsibilities associated with each functionality.
Functional categories therefore help distinguish recurring EBC proportionality from function-specific variations.

\subsection{Automated Classification of Architectural Roles}
\label{S2:AutomatedClassification}
Automated architectural-role classification assigns design responsibilities to source-code elements without requiring each element to be inspected manually.
Existing approaches mainly identify role stereotypes at method and class levels~\cite{dragan2010automatic,moreno2012jstereocode,hoquang2022role}.
Method-level labels describe individual operations, while class-level labels summarise the responsibilities expressed by a group of methods.
File-level classification uses a broader unit because a Java source file may contain several classes and methods with different responsibilities.
A file-level EBC label should therefore be interpreted as the dominant responsibility detected for the file rather than a complete description of every responsibility within it.

Early automated approaches classified architectural roles through explicit rules derived from source-code structure.
\citet{dragan2010automatic} used static analysis to identify method stereotypes, represented their frequency and distribution as a class signature, and applied empirically derived rules to infer class stereotypes.
\citet{moreno2012jstereocode} extended this approach to Java through JStereoCode, an Eclipse plug-in that automatically assigns stereotypes to methods and classes.
These rule-based approaches make the classification logic explicit and do not require a labelled training dataset, but their rules depend on predefined structural patterns and language-specific source-code analysis.
They established the feasibility of automated role classification, although their evaluations covered relatively small collections of software systems.

Data-driven classification has been introduced to reduce reliance on fixed decision rules and support more complete role assignment.
\citet{hoquang2022role} trained machine-learning models on a manually labelled dataset of 1,547 Java classes from three projects and classified the six role stereotypes defined by Wirfs-Brock.
The study showed that architectural roles can be inferred from structural source-code features, although classification performance varied across roles and decreased for less frequent stereotypes.

The present study differs in its role taxonomy, unit of classification, and analytical purpose: it assigns one EBC role to each Java source file so that aggregate role proportions can be examined across 732 projects and 78 functional categories. The resulting labels are therefore used to measure project- and category-level distributions rather than to provide a complete description of every responsibility within each source file.

\subsection{Recurring Empirical Patterns and Research Gap}
\label{S2:EmpiricalPatterns}
Software engineering research has long reported recurring empirical patterns in the structure and evolution of software systems.
\citet{lehman1980programs} formulated laws of software evolution from quantitative observations of how large software systems change over time.
\citet{wheeldon2003power} later identified repeated power-law distributions in several forms of class relationships across Java programs.
These findings show that regularities can emerge from software data even when the analysed systems differ in size and purpose.
Such regularities should be interpreted as observations within a defined empirical scope rather than as universal rules that every software system must follow.

Recurring empirical patterns are meaningful only when their scope and variation are stated clearly.
\citet{nagappan2013diversity} showed that reasoning about generality depends on the diversity and coverage of the analysed projects, not only on sample size.
Recurrence therefore does not require every observation to have identical values.
In this study, recurring EBC proportionality denotes a dominant group of functional categories with similar Entity, Boundary, and Control proportions, while categories outside that group are treated as variations that require separate interpretation.
The term observed empirical pattern therefore describes a frequent quantitative configuration within the analysed population, not a deterministic rule for all Java software.

Evidence remains limited on whether EBC roles occur in recurring proportions across a large and functionally diverse collection of Java projects.
Prior studies focused on identifying role stereotypes or examining their evolution at method and class levels using smaller sets of software systems~\cite{dragan2010automatic,moreno2012jstereocode,hoquang2022role}.
These studies did not examine how EBC proportions are distributed across functional categories, whether a dominant proportional pattern recurs across those categories, or how variations relate to category-specific architectural characteristics.
The present study addresses this gap by analyzing file-level classifier-derived EBC proportions across 732 Java projects and 78 functional categories.
This analysis connects automated architectural-role classification with the study of recurring project- and category-level patterns.

The closest empirical studies differ from the present study in their unit of analysis, empirical scale, and research purpose.
Table~\ref{tab:S2-study-comparison} provides a direct comparison.
The table includes studies that identify architectural roles or examine their distributions in source code.

\begin{table*}[htbp]
\centering
\caption{Comparison with the closest empirical studies of architectural roles.}
\label{tab:S2-study-comparison}
\small
\renewcommand{\arraystretch}{1.15}
\begin{tabularx}{\textwidth}{
@{}
>{\raggedright\arraybackslash}p{0.13\textwidth}|
*{5}{>{\raggedright\arraybackslash}X}
@{}}
\toprule
\textbf{Dimension} &
\textbf{Dragan et al.}~\cite{dragan2010automatic} &
\textbf{Moreno and Marcus}~\cite{moreno2012jstereocode} &
\textbf{Ho-Quang et al.}~\cite{hoquang2022role} &
\textbf{Fudholi and Capiluppi}~\cite{fudholi2026enhancing} &
\textbf{Present study} \\
\midrule

Language &
C++ &
Java &
Java &
Java &
Java \\

Source-code unit &
Method and class &
Method and class &
Class &
File &
File, project, and functional category \\

Empirical scope &
Multiple systems used for taxonomy construction and evaluation &
Java projects analysed using JStereoCode &
1,547 classes from three projects &
Java files used in an educational evaluation &
1,556,985 files from 732 project instances in 78 categories \\

Main purpose &
Automatic identification of class stereotypes &
Tool-supported identification of method and class stereotypes &
Classification and evolution of architectural roles &
Automatic EBC labelling and explanation &
Analysis of EBC distribution, recurrence, and variation \\

Functional-category analysis &
No &
No &
No &
No &
\textbf{Yes} \\

Recurring composition &
No &
No &
No &
No &
\textbf{Yes} \\

\bottomrule
\end{tabularx}
\end{table*}

The comparison makes the difference of the present study explicit.
Previous studies primarily developed or evaluated methods for assigning architectural roles to individual methods, classes, or files.
Some studies also examined how role distributions changed within a small number of projects.
The present study instead aggregates file-level EBC labels at project and functional-category levels and examines whether a similar composition recurs across a large software corpus.
Its main contribution is therefore the large-scale characterisation and clustering of classifier-derived EBC proportions, rather than the development of another architectural-role classifier.

\section{Methodology}
\label{S3:Methodology}

The methodology was designed to examine classifier-derived EBC proportionality across Java projects and functional categories.
Java source files served as the classification units, while projects and functional categories served as the aggregation units.
The procedure included dataset acquisition and curation, automated EBC annotation, construction of project- and category-level proportions, and analysis aligned with each research question.
The automated annotation procedure trained a classifier from weak labels and evaluated its predictions against an expert-annotated reference dataset before applying it to the full dataset.
The following subsections define the operational terms and describe each methodological stage in sufficient detail to support interpretation and replication.

\subsection{Operational Definitions}
\label{S3:Definitions}
Operational definitions specify how the main concepts were represented in the analysis.
Table~\ref{tab} defines the units, measurements, and analytical terms used throughout the study.
The architectural meanings of Entity, Boundary, and Control follow Section~\ref{S2:EBC}, while the definitions below specify how these roles were represented at file, project, and functional-category levels.

\begin{table}[htbp]
\renewcommand{\arraystretch}{1.5} 
\caption{Operational definitions used in the study.}
\label{tab}
\centering
\begin{tabular}{@{}p{0.23\linewidth}p{0.69\linewidth}@{}}
\toprule
\textbf{Term} & \textbf{Operational definition} \\
\midrule

Java source file &
A file with the \texttt{.java} extension that remained after dataset filtering and was submitted to the EBC classifier. It represents the classification unit of the study. \\

Project &
A retained Java codebase used as one analysis unit.
When two major versions of the same parent project were retained, each version was treated as a separate project instance.
\\

Functional category &
A group of projects organised according to the primary functionality assigned by the Awesome Java maintainers. It represents the second aggregation unit of the study. \\

EBC label &
One of the Entity, Boundary, or Control roles assigned to a Java source file by the classifier. The label represents the dominant responsibility detected within the file. \\

Project-level classifier-derived EBC proportion &
The number of files assigned a particular EBC role within a project divided by the total number of classified Java files in that project. \\

Category-level classifier-derived EBC proportion &
The total number of files assigned a particular EBC role across all projects in a functional category divided by the total number of classified Java files in that category. \\

Cluster &
A group of functional categories with similar Entity, Boundary, and Control proportions identified through density-based clustering. \\

Dominant cluster &
The cluster containing the largest number of functional categories using the selected DBSCAN parameters. Its distribution is used to characterised the dominant EBC compositions, derived by the classifier. \\

Dominant classifier-derived EBC composition & The average share of Entity, Boundary and Control files among the functional categories that form the largest DBSCAN cluster, identified under the  parameters in this study. This is an observed pattern, rather than a required target.\\

Non-dominant cluster &
Any identified cluster other than the dominant cluster. Non-dominant clusters represent less frequently observed EBC compositions. \\

Recurring classifier-derived EBC proportionality &
The frequently observed Entity, Boundary, and Control composition characterised from the central values and variation of the dominant cluster. \\

EBC Proportionality Law
&
The name given to the recurring classifier-derived EBC proportionality identified through RQ2.
It denotes an empirical pattern within the studied corpus rather than a universal or required architectural rule.
\\

Variation &
An EBC composition represented by a non-dominant cluster and compared with the dominant EBC compositions, derived by the classifier.  
\\

\bottomrule
\end{tabular}
\end{table}

\subsection{Study Design}
\label{S3:StudyDesign}

\begin{figure}[htbp]
    \centering
    \caption{Overview of dataset construction, automated EBC annotation,
    aggregation, and research-question-specific analysis.}
    \label{fig:methodology_overview}
    \includegraphics[
        width=\linewidth,
        height=0.82\textheight,
        keepaspectratio
    ]{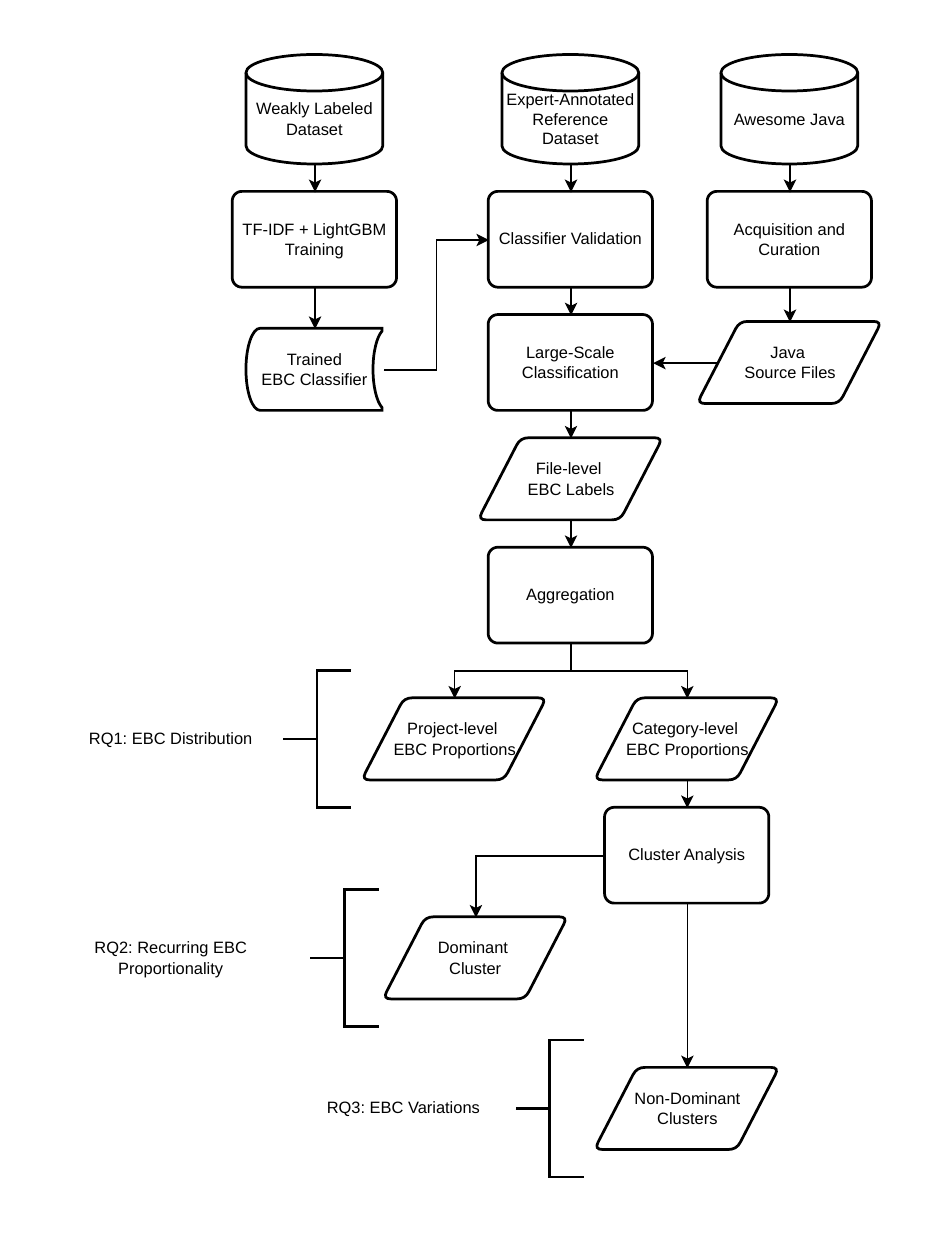}
\end{figure}

The study used a sequential empirical design to examine EBC proportionality across Java software.
Figure~\ref{fig:methodology_overview} summariseds the procedure from classifier development and dataset construction to aggregation and analysis.
The design consisted of three main parts: classifier development and expert comparison, large-scale annotation, and research-question-specific analysis.

The automated EBC annotation procedure implemented the first two parts of the study design.
It contained four stages: weak-label generation, feature representation and classifier training, comparison with expert annotations, and large-scale annotation.
These stages produced a trained classifier, provided a human reference for its predictions, and assigned EBC labels to the Java files in the main study dataset.

The datasets served different methodological purposes.
The weakly labelled dataset was represented using TF--IDF features and used to train a LightGBM classifier.
The expert-annotated reference dataset was used only to compare the trained classifier with human-assigned EBC labels.
The expert labels were not used during classifier training, although some corresponding source files were present in the weak-label training corpus.
Awesome Java~\cite{awesomejava} provided the initial collection of project candidates.
The acquired projects were curated to produce the Java source files used during large-scale annotation.

The analytical procedure separated file-level classification from project- and category-level analysis.
After classifier development and expert comparison, each retained Java source file was assigned one EBC label.
The file-level labels were then aggregated into project-level and category-level classifier-derived EBC proportions.
Both aggregation levels were used to answer RQ1 by describing how classifier-derived EBC proportionality was distributed across projects and software functional categories.

The cluster analysis separated recurring classifier-derived EBC proportionality from variations in that proportionality.
Category-level classifier-derived EBC proportions were used as the input to the cluster analysis.
The largest cluster, based on the number of functional categories, was defined as the dominant cluster and used to characterised the recurring proportionality for RQ2.
The remaining clusters were treated as non-dominant clusters and used to identify variations for RQ3.
The quantitative analysis covered all non-dominant clusters, while the detailed interpretation focused on representative functional categories to keep the presentation concise.
The interpretation examined whether the observed variations were consistent with the architectural characteristics of those functional categories.

\subsection{Dataset Acquisition and Curation}
\label{S3:Dataset}

The dataset acquisition procedure constructed the study dataset from a curated collection of Java software.
Awesome Java~\cite{awesomejava}, a public collection of Java frameworks, libraries, and software, was used as the initial data source.
The collection was retrieved in April 2026 and contained 830 project entries across 83 functional categories before filtering.
The categories provided by Awesome Java were retained as the functional-category labels used in the study.

The data source was selected to provide functional diversity across Java software.
Awesome Java covers categories such as Artificial Intelligence, Messaging, Security, Database, Web Frameworks, and Testing.
\citet{sas2021using} also used Awesome Java as a source of Java projects in an empirical software engineering study.
However, the resulting dataset is treated as a broad curated sample rather than a random sample of the complete Java ecosystem.

The curation procedure applied three exclusion criteria.
First, an entry was excluded when its source-code repository had been deleted, made private, become inaccessible, or moved to a closed-source model.
Second, non-project entries, including forums, documentation websites, learning resources, and references to individuals, were excluded.
Third, an acquired repository was excluded when it did not contain any \texttt{.java} files.
These criteria were applied before EBC annotation and did not depend on the EBC distribution of a project.

The filtering process produced the final set of projects used in the study.
One retained project provided two major versions, and each version was analysed as a separate project instance within the same functional category.
The final dataset therefore contained 732 project instances across 78 functional categories.
Table~\ref{tab:dataset_curation} summariseds the acquisition process, the applied exclusions, and the final dataset composition.

\begin{table}[htbp]
\centering
\caption{Dataset acquisition, filtering, and final composition.}
\label{tab:dataset_curation}
\begin{tabular}{lr}
\toprule
\textbf{Dataset item} & \textbf{Count} \\
\midrule
Initial functional categories                  & 83 \\
Initial project entries                        & 830 \\
\midrule
Unavailable repositories removed               & 32 \\
Non-project entries removed                    & 45 \\
Repositories without Java files removed        & 22 \\
Major-version instance added                   & 1 \\
\midrule
Final project instances                        & 732 \\
Final functional categories                    & 78 \\
Paths ending in \texttt{.java}              & 1,556,998 \\
Classified Java files                       & 1,556,985 \\
\bottomrule
\end{tabular}
\end{table}

The curated dataset provided the Java source files used for automated EBC annotation.
It contained 732 project instances distributed across 78 functional categories.
The complete project list, category assignments, acquisition metadata, analysis scripts, and generated results are available in the Zenodo replication package.\footnote{\url{https://doi.org/10.5281/zenodo.22746260}}
The file \texttt{Awesome Java Projects List and Documentation} provides the main guide to the package contents.

\subsection{Weak-Label Generation}
\label{S3:WeakLabels}

Weak-label generation provided the training targets for the file-level EBC classifier.
Weak supervision supports the construction of training datasets when complete manual annotation is impractical~\cite{ratner2017snorkel}.
Previous research has applied weak labelling to file-level source-code classification~\cite{sas2023weaklabelling}, while recent work has used large language models to generate labels for downstream classifiers~\cite{pangakis2024knowledge}.
These approaches provide the methodological basis for constructing the EBC training corpus.

The weak-label corpus was constructed independently of the main analysis dataset.
Nine open-source Java projects were purposively selected because they contained many Java source files.
The projects were Mindustry, SmartTube, agentscope-java, arthas, baritone, jabref, pumpernickel, spring-ai-alibaba, and strykerapp.
None of these projects overlapped with the 732 project instances used to answer the research questions.
Together, the nine projects contributed 10,272 Java files for weak labelling.

Each Java file was processed as one annotation unit.
The request contained the relative file path and the complete source text without client-side truncation.
The \texttt{gpt-5-mini} model identifier was invoked through the OpenAI Chat Completions API~\cite{openai2026gpt5mini}.
The requests were submitted between February 23 and March 4, 2026.
No optional reasoning, sampling, or output-token parameters were explicitly set.
The API defaults available during the annotation period were therefore used.

The annotation prompt and response schema defined the information returned for each file.
The prompt described Boundary as an external interface, Control as workflow coordination, and Entity as domain data or state.
It requested three EBC entries with their probabilities, supporting code evidence, and a short explanation.
The response schema restricted the EBC labels to Boundary, Control, and Entity and restricted each probability to the interval from zero to one.
The same response contained auxiliary software-topic information based on the 301-label GitRanking taxonomy~\cite{sas2023gitranking}.
All auxiliary topic information was excluded from the present study; only the first returned EBC label and its associated confidence were carried forward to classifier construction.

The first returned EBC role was treated as the representative EBC label of a file.
This operational rule provided one training target for each file while allowing the file to contain evidence of more than one architectural responsibility.
The probability associated with the representative label was treated as label confidence and was later used to weight the training instance.
The representative label therefore identified the role selected for classification without asserting that the file contained only one responsibility.

The generated responses were checked before the training corpus was constructed.
The checks considered the expected response length, probability range, approximate probability sum, and presence of an explanation.
The parsed values, warning flags, model identifier, prompt version, annotation time, and raw structured response were retained for inspection.
The complete prompt, response schema, and auxiliary topic-label list are provided in the replication package.

The review indicator recorded possible uncertainty without automatically excluding an annotation.
It considered low topic confidence, low EBC confidence, response warnings, and model-generated flags such as \texttt{MIXED\_ROLE}.
Because the indicator also covered the auxiliary topic task and files with possible combinations of responsibilities, it was retained as audit information rather than used as an exclusion criterion.
Failed requests were recorded, and processing continued with the remaining files.
The dataset was saved after every request, although no automatic retry mechanism was applied within a run.

The final weak-label corpus represented the output of the annotation and filtering procedure.
One API request failed because of a connection error and did not produce an EBC label, leaving 10,271 successfully labelled files.
During training preparation, one \texttt{package-info.java} file containing only a package declaration was removed because its source text contained fewer than 31 characters.
Table~\ref{tab:weak_label_dataset} presents the representative-label composition of the resulting 10,270-file training corpus.
Control was the most frequent representative label, followed by Boundary and Entity.
This corpus formed the input to the feature-representation and classifier-training procedure described next.

\begin{table}[htbp]
\centering
\caption{Composition of the weak-label training corpus by representative EBC label.}
\label{tab:weak_label_dataset}
\begin{tabular}{lrr}
\toprule
\textbf{Representative label} & \textbf{Files} & \textbf{Percentage} \\
\midrule
Entity   & 1,780 & 17.33\% \\
Boundary & 4,054 & 39.47\% \\
Control  & 4,436 & 43.19\% \\
\midrule
\textbf{Total} & \textbf{10,270} & \textbf{100.00\%} \\
\bottomrule
\end{tabular}
\end{table}

\subsection{Feature Representation and Classifier Training}
\label{S3:Training}

The classifier-training procedure transformed each Java file into a numerical representation and learned its representative EBC label.
Term frequency--inverse document frequency (TF--IDF) was used to represent the source text~\cite{salton1988term}.
LightGBM was used to learn the relationship between the resulting features and the EBC labels~\cite{ke2017lightgbm}.
Both components were implemented as a single scikit-learn pipeline~\cite{pedregosa2011scikit}.
This pipeline ensured that the same feature transformation was applied during training and later annotation.

The training input consisted of the 10,270 files retained from the weak-label corpus.
The complete source text was used as the classifier input, and the representative EBC label was used as the training target.
Comments, string literals, and other source-code content were not removed from the final representation.
The files were divided using a stratified file-level split, with 80\% used for training and 20\% used for internal assessment.
A random seed of 42 was used to make the split reproducible.
The split produced 8,216 training files and 2,054 held-out files while preserving the representative-label distribution.

The feature representation captured individual source-code tokens and adjacent token pairs.
The TF--IDF vectorisedr used unigrams and bigrams and retained at most 50,000 features.
The token pattern \verb|\b[a-zA-Z_]\w+\b| retained tokens beginning with a letter or underscore.
The remaining vectoriser options used the scikit-learn defaults, including lowercase conversion, smoothed inverse document frequency, and Euclidean normalisation.
The vectoriser was fitted only on the training portion of the corpus.

The label confidence controlled the contribution of each training file.
For file \(i\), the probability \(p_i\) represents the confidence associated with its representative EBC label, while \(w_i\) represents its training weight.
Equation~\ref{eq:weak_label_weight} defines the relationship between these values.
\begin{equation}
w_i = \max\left(0.05,\,2(p_i - 0.5)\right).
\label{eq:weak_label_weight}
\end{equation}
Higher-confidence labels received greater influence during classifier training.
The minimum weight of 0.05 retained low-confidence annotations while limiting their influence on the model.

The LightGBM classifier learned a three-class mapping from the TF--IDF vectors to Boundary, Control, and Entity.
The classifier used 300 boosting iterations, a learning rate of 0.1, 31 leaves per tree, and a random seed of 42.
Training used all available processor cores, while the remaining LightGBM parameters used their library defaults.
No additional class-balancing weights, resampling procedure, or hyperparameter search was applied.
The confidence weights in Equation~\ref{eq:weak_label_weight} were passed to LightGBM during model fitting.

The held-out files provided an internal assessment of agreement with the generated weak labels.
The confidence-weighted model achieved 72.44\% accuracy and a macro-F1 score of 0.72 on the 2,054 held-out files.
These measurements describe how consistently the trained classifier reproduced the weak labels and do not constitute independent validation of the EBC classifications.
The trained pipeline was stored as the final file-level EBC classifier and evaluated against expert annotations in the following subsection.

\subsection{Comparison with Expert Annotations}
\label{S3:ExpertComparison}

The expert comparison provided a human reference for assessing file-level EBC predictions.
The reference annotations were originally constructed for an earlier study of automatic EBC labelling and explanation~\cite{fudholi2026enhancing}.
That study also used explainable artificial intelligence (XAI) to examine the information supporting the predicted labels.
The present study reused only the file-level EBC labels from the reference dataset.
The explanation scores and other XAI outputs were excluded because the present analysis focuses on classifier-derived EBC proportionality rather than individual prediction explanations.

The reference set was constructed through folder-level sampling.
Two source folders within the Pumpernickel project, \texttt{com/pump/awt} and \texttt{com/pump/data}, were selected at random, after which the Java files within these folders and their subfolders were included.
This procedure produced 102 Java files across ten Java packages.
The reference set contained 53 Entity files, 27 Control files, and 22 Boundary files.
The random selection therefore occurred at the folder level and did not constitute a random sample of individual files from the complete project.

The reference labels were produced through an independent expert-annotation process.
The second author first assigned one representative EBC role to each file independently.
The earlier study reports that the annotator had more than 20 years of software-engineering experience as both a practitioner and an educator~\cite{fudholi2026enhancing}.
The labels were created before the predictions examined in the present study and therefore were not influenced by the current classifier.

The comparison set was not fully independent of classifier construction.
The same 102 source files were also present in the Pumpernickel weak-label corpus, although their expert labels were not used for classifier training.
The file-level split placed 77 of these files in the fitted training portion and 25 files in the held-out portion.
Consequently, the complete set was used as a diagnostic comparison with expert labels rather than as an independent test set.
Results for the 25 files that were not used for fitting were also calculated separately.

The comparison measured accuracy, class-level precision, recall, and F1-score, as well as the macro-averaged F1-score and the confusion matrix~\cite{sokolova2009systematic}.
A majority-label baseline was included because the expert-labelled set contained unequal numbers of Entity, Boundary, and Control files.
These measures were used to identify agreement patterns and class-specific weaknesses in the automated annotations.

\subsection{Large-Scale EBC Annotation}
\label{S3:LargeScaleAnnotation}

Large-scale annotation assigned one representative EBC role to each readable Java file in the final dataset.
The process used the trained TF--IDF and LightGBM pipeline described in the preceding subsections.
Each Java file remained one annotation unit, following the file-level classification approach used in previous software-labelling research~\cite{sas2023weaklabelling}.
This stage converted the curated source-code collection into file-level EBC predictions.

The source files were collected through recursive scanning within each retained project.
Every path ending in \texttt{.java} was considered, without applying separate path-based filters for production, test, example, or generated-source directories.
Each readable file was decoded as UTF-8, with invalid characters ignored, and its complete text was passed to the classifier without truncation.
Thirteen matched paths were directories whose names ended in \texttt{.java} and were excluded because they could not be read as source files.

The prediction procedure used the fixed model produced during classifier training.
For each source file, the saved pipeline applied the fitted TF--IDF representation and calculated the probabilities of the Entity, Boundary, and Control classes using LightGBM~\cite{pedregosa2011scikit,ke2017lightgbm}.
The three roles were ordered by their predicted probabilities, and the role with the highest probability was stored as the representative EBC label.
No minimum-confidence threshold was applied, while the probabilities of all three roles were retained for inspection.

The annotation process was performed one project at a time.
Results were buffered and saved after every 100 processed files, while project-level progress records allowed an interrupted run to continue from previously recorded paths.
Each output record contained the functional category, project name, file path, file name, package name, file length, processing status, predicted roles, and their probabilities.
The source text was not duplicated in the output files.

The large-scale annotation produced the file-level dataset used in the subsequent analysis.
Valid predictions were obtained for 1,556,985 Java files from 732 project instances across 78 functional categories.
The resulting representative labels were then used to construct the project-level classifier-derived EBC proportions described in the next subsection.

\subsection{Construction of EBC Proportions}
\label{S3:Proportions}

The construction of classifier-derived EBC proportions transformed the file-level predictions into comparable project and functional-category compositions.
For each project instance \(p\), the numbers of files assigned to Entity, Boundary, and Control were denoted by \(n_{p,E}\), \(n_{p,B}\), and \(n_{p,C}\), respectively.
Equation~\ref{eq:project-total} defines the total number of classified files \(N_p\) in project \(p\).
\begin{equation}
N_p = n_{p,E} + n_{p,B} + n_{p,C}.
\label{eq:project-total}
\end{equation}

Project-level classifier-derived EBC proportionality describes the composition of an individual project.
For each role \(r \in \{E,B,C\}\), Equation~\ref{eq:project-proportion} defines its project-level proportion \(P_{p,r}\).
\begin{equation}
P_{p,r} = \frac{n_{p,r}}{N_p}.
\label{eq:project-proportion}
\end{equation}

The three project-level proportions ranged from zero to one and summed to one for every project.
They therefore formed compositional observations in which the three components represented parts of the same whole~\cite{aitchison1986statistical}.
The proportions were multiplied by 100 when reported as percentages in tables and figures.

Functional-category classifier-derived EBC proportionality describes the combined file composition of the projects assigned to the same category.
For each functional category \(g\), Equation~\ref{eq:category-role-count} defines the total number of files \(n_{g,r}\) assigned to role \(r\).
\begin{equation}
n_{g,r} = \sum_{p \in g} n_{p,r}.
\label{eq:category-role-count}
\end{equation}

The functional-category proportion represents the contribution of each EBC role to the complete category.
Equation~\ref{eq:category-proportion} defines the proportion \(P_{g,r}\) of role \(r\) in category \(g\).
\begin{equation}
P_{g,r} =
\frac{n_{g,r}}
{n_{g,E} + n_{g,B} + n_{g,C}}.
\label{eq:category-proportion}
\end{equation}

This calculation pooled the classified files of all projects within a functional category.
It did not calculate the arithmetic mean of the project-level proportions.
Each successfully classified file therefore contributed equally to the category composition, while projects containing more Java files contributed more observations.

The final proportions included only files with a valid EBC label.
The 13 paths that could not be processed during large-scale annotation were excluded, leaving 1,556,985 classified files.
A role count was recorded as zero when no file in a project was assigned to that role.
The project remained in the analysis when at least one valid prediction was available.
The resulting project-level and functional-category-level proportions formed the inputs to the analyses for the three research questions.
\subsection{Analysis by Research Question}
\label{S3:Analysis}
The analysis was organised according to the three research questions.
RQ1 described the distribution of classifier-derived EBC proportions at project and functional-category levels.
RQ2 identified recurring patterns among the functional-category proportions.
RQ3 examined the functional categories that varied from the recurring pattern.
Each analysis used the proportions defined in Section~\ref{S3:Proportions}.

\subsubsection{RQ1: Distribution Across Projects and Functional Categories}
The analysis for RQ1 described how classifier-derived EBC proportionality was distributed across the dataset.
The project-level analysis used the Entity, Boundary, and Control proportions of the 732 project instances.
The functional-category-level analysis used the corresponding pooled proportions of the 78 functional categories.
The two analytical levels were examined separately.

The distribution of each EBC role was summarised using the mean, standard deviation, minimum, first quartile, median, third quartile, and maximum.
The interquartile range was calculated as the difference between the third and first quartiles.
These statistics described both the central proportions and the amount of variation across projects and functional categories.

The distributions were also examined using boxplots.
Boxplots provide a compact representation of the median, interquartile range, and observations beyond the central distribution~\cite{tukey1977exploratory}.
Separate boxplots were produced for the project and functional-category levels so that differences in their dispersion could be compared directly.

The RQ1 analysis was descriptive and did not apply a statistical significance test.
Its purpose was to establish the observed classifier-derived EBC proportions and their variation rather than to test a directional hypothesis.
The resulting statistics provided the empirical basis for identifying recurring proportionality in RQ2.

\subsubsection{RQ2: Identification of Recurring classifier-derived EBC proportionality}

The RQ2 analysis examined whether similar classifier-derived EBC proportions recurred across functional categories.
Each of the 78 functional categories was represented by its pooled Entity, Boundary, and Control proportions.
Before clustering, the three proportions were standardised to a mean of zero and a standard deviation of one using \texttt{StandardScaler}~\cite{pedregosa2011scikit}.
This transformation placed the three EBC components on comparable scales during clustering.

The standardised category proportions were grouped using the DBSCAN algorithm.
DBSCAN can identify groups of nearby observations without requiring the number of clusters to be specified in advance~\cite{ester1996density}.
Euclidean distance was used to measure the distance between functional categories.
The \texttt{min\_samples} parameter was set to one so that every functional category remained available for the later analysis of variations.
Under this setting, the procedure formed groups through connected neighborhoods and did not classify any category as noise.

The neighborhood size was examined across several parameter values.
The value of (\(\varepsilon\)) was varied from (0.2) to (0.8) in increments of (0.1).
The number of clusters and the silhouette coefficient were recorded for each value~\cite{rousseeuw1987silhouettes}.
The final analysis used (\(\varepsilon\) = 0.6), which provided an intermediate partition between the fragmented groups produced by smaller values and the broad merging produced by larger values.
The silhouette coefficient was used as a diagnostic measure rather than as the sole parameter-selection criterion.

The recurring EBC pattern was identified from the largest cluster in the final partition.
Its coverage was measured using the numbers and percentages of functional categories, project instances, and classified Java files represented by its members.
Its central composition was calculated as the arithmetic mean of the category-level classifier-derived EBC proportions, giving equal weight to every functional category.
The median, standard deviation, quartiles, and interquartile range were also calculated for each EBC role to describe the variation within the cluster.
A cluster with broad coverage and concentrated classifier-derived EBC proportions was treated as evidence of a recurring empirical pattern rather than a universal architectural rule.

The cluster structure was presented using a ternary plot of the original category-level proportions.
The plot shows the relative Entity, Boundary, and Control composition of each functional category and indicates its assigned cluster.
The plot was used for presentation only, while the cluster assignments were obtained from the standardised proportions.

\subsubsection{RQ3: Analysis of Variations}

The RQ3 analysis examined how functional categories differed from the recurring classifier-derived EBC proportionality.
Every functional category outside the largest cluster was treated as a variation from the recurring pattern.
For each category, its Entity, Boundary, and Control proportions were compared with the corresponding distributions within the recurring cluster.
This comparison identified the direction and magnitude of each variation.

The variation levels were defined using the quartiles and interquartile ranges of the recurring cluster.
A proportion between the first and third quartiles was classified as \textsc{Typical}.
A proportion between the lower IQR bound and the first quartile was classified as \textsc{Low}, while a proportion below the lower bound was classified as \textsc{Very Low}.
A proportion between the third quartile and the upper IQR bound was classified as \textsc{High}, while a proportion above the upper bound was classified as \textsc{Very High}.
The lower and upper bounds were calculated by subtracting and adding 1.5 times the IQR from the first and third quartiles, respectively~\cite{tukey1977exploratory}.
These labels described differences from the recurring pattern and did not indicate architectural quality or correctness.

The non-dominant clusters were examined according to their coverage and EBC composition.
The largest non-dominant cluster was treated as an alternative proportionality pattern because it represented substantially more functional categories than the other non-dominant clusters.
The remaining clusters were treated as specialised proportionality patterns because they represented smaller groups of functional categories.
Their compositions were compared with the recurring cluster to identify Entity-oriented, Boundary-oriented, Control-oriented, and combined variations.

The architectural assessment examined selected representative variations.
The selected cases covered different EBC roles and different directions of variation and were supported by published descriptions of the corresponding software functionality.
The reported architectural responsibilities were compared with the operational definitions of Entity, Boundary, and Control used in this study.
Evidence was considered consistent when the responsibilities associated with a functional category agreed with the direction of its observed variation.
Evidence was considered less consistent when the literature described responsibilities that suggested a stronger presence of an EBC role than the observed proportion indicated.

The architectural assessment was performed separately for each EBC role.
A functional category could therefore contain evidence that supported one part of its proportionality while questioning another part.
Such cases were treated as mixed rather than assigning a single judgment to the complete functional category.
The selected cases provided explanatory examples and did not constitute an exhaustive validation of every functional category outside the recurring cluster.

\section{Results}
\label{S4:Result}

This section presents the empirical results obtained from the final classified dataset.
The first subsection describes the analytical corpus and the coverage of the automated annotation.
The second subsection reports the comparison between the classifier predictions and the expert annotations.
The remaining subsections answer the three research questions.
RQ1 describes the distribution of classifier-derived EBC proportionality across projects and functional categories.
RQ2 examines the recurring classifier-derived EBC proportionality identified across the functional categories.
RQ3 describes the variations from this recurring pattern and examines selected variations in relation to their architectural characteristics.

\subsection{Analytical Corpus and Annotation Coverage}
\label{S4:Corpus}

The analytical corpus contained the observations used to answer the three research questions.
The final dataset contained 732 project instances representing 731 parent projects across 78 functional categories.
The annotation scan identified 1,556,998 paths ending in \texttt{.java}.
Thirteen of these paths were directories rather than readable source files and were excluded.
The resulting analytical corpus contained 1,556,985 Java files with valid EBC predictions.
Table~\ref{tab:corpus} summarises the composition and annotation coverage of the corpus.

\begin{table}[htbp]
\centering
\caption{Composition and annotation coverage of the analytical corpus.}
\label{tab:corpus}
\begin{tabular}{lr}
\toprule
\textbf{Corpus item} & \textbf{Count} \\
\midrule
Parent projects                         & 731 \\
Project instances                       & 732 \\
Functional categories                   & 78 \\
Paths ending in \texttt{.java}          & 1,556,998 \\
Non-file paths excluded                 & 13 \\
Classified Java files                   & 1,556,985 \\
\midrule
Total LOC                               & 292,451,670 \\
Non-blank LOC                           & 258,876,768 \\
Approximate code LOC                    & 174,680,537 \\
\bottomrule
\end{tabular}
\end{table}

The LOC measurements describe different parts of the source text.
Total LOC includes every line in the readable Java files, while non-blank LOC excludes empty lines.
Approximate code LOC additionally excludes lines identified as comments using the line-based counting procedure.
This final measurement is used only as a descriptive estimate because line-based counting cannot identify every multiline or mixed code-and-comment case.

The project instances covered a broad range of software sizes.
Table~\ref{tab:project-size-statistics} presents the distributions of classified Java files and LOC across the 732 project instances.
The median project contained 392 classified Java files and 54,354.50 total LOC, whereas the corresponding means were 2,127.03 files and 399,524.14 LOC.
The differences between the median and mean indicate a right-skewed distribution containing many small and medium-sised projects together with several substantially larger projects.

\begin{table}[htbp]
\centering
\caption{Project size statistics across the analysed repositories.}
\label{tab:project-size-statistics}
\begin{tabular}{l|rrrrr|r}
\toprule
Metric & Min & Q1 & Median & Q3 & Max & Mean \\
\midrule
Java files & 1 & 113.00 & 392.00 & 1,187.00 & 153,188 & 2,127.03 \\
Total LOC & 8 & 14,155.25 & 54,354.50 & 182,741.75 & 57,876,256 & 399,524.14 \\
Non-blank LOC & 6 & 12,151.50 & 46,985.50 & 157,925.75 & 53,310,549 & 353,656.79 \\
Approximate code LOC & 3 & 8,757.50 & 32,565.00 & 112,161.75 & 31,375,336 & 238,634.61 \\
\bottomrule
\end{tabular}
\end{table}

The functional categories differed in their numbers of represented projects.
Table~\ref{tab:top-categories} presents the ten largest functional categories.
The largest category, Platform--Apache Commons, contained 47 project instances and represented 6.42\% of the corpus.
The ten largest categories together represented 35.52\% of the project instances, while the remaining categories represented 64.48\%.
A functional category contained a median of seven project instances, and only two categories contained a single project instance.
The analytical corpus was therefore not dominated by one functional category or by a small group of categories.

\begin{table}[htbp]
\centering
\caption{Top ten functional categories by number of projects.}
\label{tab:top-categories}
\begin{tabular}{lrr}
\toprule
Functional category & Projects & Percentage \\
\midrule
Platform--Apache Commons & 47 & 6.42\% \\
Database & 39 & 5.33\% \\
Miscellaneous & 37 & 5.05\% \\
Monitoring & 25 & 3.42\% \\
Utility & 24 & 3.28\% \\
Security & 24 & 3.28\% \\
Web Frameworks & 17 & 2.32\% \\
JSON & 16 & 2.19\% \\
Messaging & 16 & 2.19\% \\
Machine Learning & 15 & 2.05\% \\
\bottomrule
\end{tabular}
\end{table}

The corpus contained substantial variation in both project scale and assigned software functionality.
This variation provides the empirical context for examining whether similar EBC proportions occur across projects of different sizes and across different functional categories.

\subsection{Comparison with Expert Annotations}
\label{S4:ExpertComparison}

The expert comparison assessed agreement between the classifier predictions and manually assigned EBC labels.
The reference set contained 102 Java files from two randomly selected folders of Pumpernickel and was previously used to evaluate automated EBC classification~\cite{fudholi2026enhancing}.
The second author assigned the reference labels independently, and these labels were not used during classifier training.
However, Pumpernickel was included in the weak-label corpus.
The comparison was therefore treated as an expert-agreement check rather than a project-independent test.

The classifier showed limited agreement with the expert annotations.
Table~\ref{tab:S4-expert-performance} presents its performance across the 102 files.
The classifier achieved an accuracy of 31.37\%, a macro-averaged F1-score of 0.307, and a weighted F1-score of 0.288.

\begin{table}[htbp]
\centering
\caption{Classifier performance against the expert annotations.}
\label{tab:S4-expert-performance}
\begin{tabular}{lrrrr}
\toprule
\textbf{Class} &
\textbf{Precision} &
\textbf{Recall} &
\textbf{F1-score} &
\textbf{Support} \\
\midrule
Boundary & 0.20 & 0.41 & 0.27 & 22 \\
Control  & 0.33 & 0.59 & 0.43 & 27 \\
Entity   & 0.78 & 0.13 & 0.23 & 53 \\
\midrule
Accuracy &
\multicolumn{3}{r}{0.31} &
102 \\
Macro average    & 0.44 & 0.38 & 0.31 & 102 \\
Weighted average & 0.54 & 0.31 & 0.29 & 102 \\
\bottomrule
\end{tabular}
\end{table}

The classification performance differed across the three EBC roles.
Control achieved the highest recall at 59.26\%, while Entity achieved the highest precision at 77.78\%.
However, Entity recall was only 13.21\%.
The classifier therefore identified only a small proportion of the files assigned an Entity label by the expert.

The main disagreement involved expert-labelled Entity files.
Table~\ref{tab:S4-expert-confusion} presents the confusion matrix.
Only 7 of the 53 expert-labelled Entity files were predicted as Entity.
The remaining Entity files were predicted mainly as Boundary or Control.

\begin{table}[htbp]
\centering
\caption{Confusion matrix between the expert labels and classifier predictions.}
\label{tab:S4-expert-confusion}
\begin{tabular}{lrrr}
\toprule
& \multicolumn{3}{c}{\textbf{Classifier prediction}} \\
\cmidrule(lr){2-4}
\textbf{Expert label} &
\textbf{Boundary} &
\textbf{Control} &
\textbf{Entity} \\
\midrule
Boundary & 9  & 13 & 0 \\
Control  & 9  & 16 & 2 \\
Entity   & 27 & 19 & 7 \\
\bottomrule
\end{tabular}
\end{table}

The majority-class baseline provides additional context for the observed performance.
Entity represented 53 of the 102 expert labels.
An Entity-only classifier would therefore achieve 51.96\% accuracy, a macro-averaged F1-score of 0.228, and a weighted F1-score of 0.355.
The EBC classifier achieved lower accuracy and weighted F1-score but a higher macro-averaged F1-score because it produced predictions for all three roles.
The comparison does not support strong file-level classification performance.

The files not used during model fitting provided a more conservative comparison.
Across these 25 files, the classifier achieved 44.00\% accuracy, a macro-averaged F1-score of 0.426, and a weighted F1-score of 0.408.
The Entity-only baseline also achieved 44.00\% accuracy but produced a macro-averaged F1-score of 0.204 and a weighted F1-score of 0.269.
The small subset prevents a strong performance conclusion, but it shows that the classifier represented the three roles more evenly than the baseline.

The error pattern introduces uncertainty into the automated EBC measurements.
A Java file may contain characteristics associated with several EBC roles, while the classification procedure assigns only one representative label.
The comparison cannot determine whether each disagreement results from role overlap, classifier error, or differences in role interpretation.
The low Entity recall may also cause Entity proportions to be underestimated and Boundary or Control proportions to be overestimated.

The remaining analyses characterise proportionality in the classifier-assigned EBC roles.
The reported proportions should therefore be interpreted as classifier-derived measurements rather than verified architectural-role proportions.

\subsection{RQ1: Distribution Across Projects and Functional Categories}
\label{S4:RQ1}

\begin{figure*}[htbp]
\centering
\includegraphics[width=\textwidth]{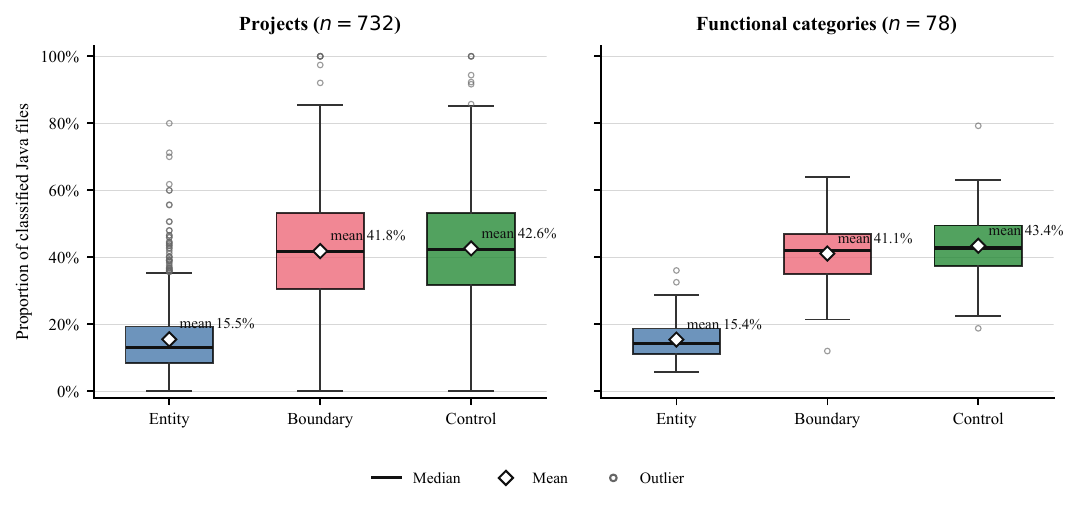}
\caption{Distribution of classifier-assigned Entity, Boundary, and Control proportions at project and functional-category levels. Boxes cover the first to third quartiles, central lines show the medians, diamonds show the arithmetic means, whiskers extend to the most extreme observations within 1.5 times the interquartile range, and circles show outliers.}
\Description{Two box-and-whisker panels compare Entity, Boundary, and Control proportions across 732 project instances and 78 functional categories. Project-level proportions show greater dispersion and more outliers than functional-category proportions.}
\label{fig:S4-ebc-distribution}
\end{figure*}

RQ1 examines classifier-derived EBC proportionality at the project and functional-category levels.
The project-level analysis covers 732 project instances, while the functional-category analysis covers 78 categories.
Table~\ref{tab:S4-ebc-distribution} reports the descriptive statistics at both levels.
Figure~\ref{fig:S4-ebc-distribution} shows the corresponding distributions.

\begin{table*}[htbp]
\centering
\caption{Distribution of classifier-derived EBC proportions across projects and functional categories. All values are percentages.}
\label{tab:S4-ebc-distribution}
\small
\setlength{\tabcolsep}{5pt}
\begin{tabular}{llrrrrrrr}
\toprule
\textbf{Level} &
\textbf{Role} &
\textbf{Mean} &
\textbf{SD} &
\textbf{Q1} &
\textbf{Median} &
\textbf{Q3} &
\textbf{Min} &
\textbf{Max} \\
\midrule
Project & Entity   & 15.51 & 10.96 & 8.54  & 13.04 & 19.34 & 0.00 & 80.00 \\
Project & Boundary & 41.85 & 16.63 & 30.63 & 41.75 & 53.33 & 0.00 & 100.00 \\
Project & Control  & 42.64 & 16.21 & 31.74 & 42.39 & 53.29 & 0.00 & 100.00 \\
\midrule
Functional category & Entity   & 15.44 & 5.63  & 11.16 & 14.23 & 18.74 & 5.88  & 36.07 \\
Functional category & Boundary & 41.12 & 10.02 & 35.11 & 41.92 & 46.96 & 12.01 & 63.97 \\
Functional category & Control  & 43.44 & 10.26 & 37.32 & 42.76 & 49.51 & 18.80 & 79.26 \\
\bottomrule
\end{tabular}
\end{table*}

The project-level distribution contains a small Entity proportion and similar Boundary and Control proportions.
The median project consists of 13.04\% Entity, 41.75\% Boundary, and 42.39\% Control files.
The corresponding means are 15.51\%, 41.85\%, and 42.64\%.
Both measures show that Entity forms the smallest proportion, while Boundary and Control form larger and comparable proportions.

The project-level proportions also vary considerably.
Entity ranges from 0\% to 80\%, while Boundary and Control both range from 0\% to 100\%.
The standard deviations are 10.96 percentage points for Entity, 16.63 percentage points for Boundary, and 16.21 percentage points for Control.
These values show that individual projects do not have an identical EBC composition.

The functional-category distribution has similar central values.
The median functional category consists of 14.23\% Entity, 41.92\% Boundary, and 42.76\% Control files.
The corresponding means are 15.44\%, 41.12\%, and 43.44\%.
Entity therefore remains the smallest proportion, while Boundary and Control remain comparable after projects are grouped by functional category.

The functional-category proportions still vary across software functionalities.
Entity ranges from 5.88\% to 36.07\%, Boundary ranges from 12.01\% to 63.97\%, and Control ranges from 18.80\% to 79.26\%.
These ranges show that some functional categories have substantially different EBC compositions.

The classifier-derived EBC proportions show different levels of dispersion at the two aggregation levels.
Figure~\ref{fig:S4-ebc-distribution} presents the distributions across project instances and functional categories.
The project-level distributions are wider and contain more outliers, particularly for Entity.
The functional-category distributions are more concentrated because each category combines the files of several projects.
The mean and median remain similar for Boundary and Control at both aggregation levels, while Entity has a lower central proportion.

\begin{center}
\fbox{
\parbox{0.94\linewidth}{
\textbf{Answer to RQ1.}
Classifier-derived EBC proportions have a similar central distribution at the project and functional-category levels.
Entity accounts for approximately 13--15\% of Java files, while Boundary and Control each account for approximately 41--43\%.
Individual projects show substantial variation: functional categories show lower dispersion while retaining differences among software functionalities.
}}
\end{center}

\subsection{RQ2: Recurring classifier-derived EBC proportionality}
\label{S4:RQ2}

\begin{figure*}[htbp]
\centering
\includegraphics[width=0.82\textwidth]{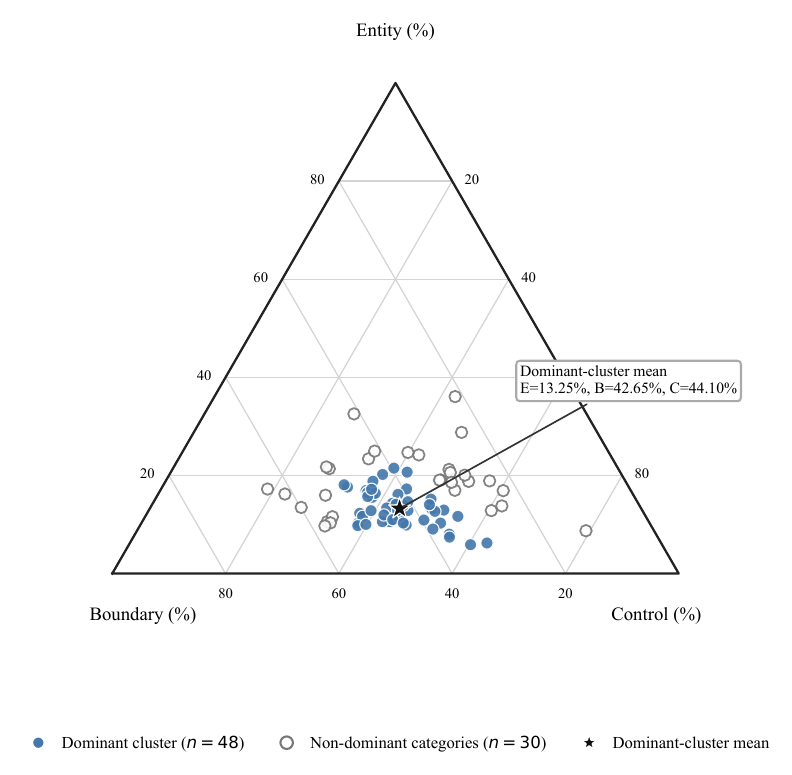}
\caption{Distribution of the 78 functional categories in the classifier-derived EBC proportionality space. Filled blue circles represent the 48 categories in the dominant cluster, open grey circles represent the 30 non-dominant categories, and the star marks the mean composition of the dominant cluster.}
\Description{A ternary plot showing Entity, Boundary, and Control proportions for 78 functional categories. The 48 categories in the dominant cluster form a concentrated region around a mean of 13.25 percent Entity, 42.65 percent Boundary, and 44.10 percent Control. The 30 non-dominant categories are more widely dispersed around this region.}
\label{fig:S4-cluster-ternary}
\end{figure*}

RQ2 examines whether similar \textbf{EBC proportions} concentrate across functional categories.
The analysis applied DBSCAN to the classifier-derived EBC proportions of the 78 functional categories.
The resulting partition contained 13 clusters.
Table~\ref{tab:S4-cluster-summary} reports the coverage and mean composition of each cluster, while Figure~\ref{fig:S4-cluster-ternary} shows their positions in the classifier-derived EBC proportionality space.

\begin{table*}[htbp]
\centering
\caption{Coverage and mean EBC composition of the functional-category clusters.}
\label{tab:S4-cluster-summary}
\small
\begin{tabular}{c|ccc|ccc}
\toprule
\textbf{Cluster} &
\textbf{Categories} &
\textbf{Projects} &
\textbf{Files} &
\textbf{Entity} &
\textbf{Boundary} &
\textbf{Control} \\
\midrule
\textbf{0} & \textbf{48} & \textbf{472} & \textbf{743,407} & \textbf{13.25\%} & \textbf{42.65\%} & \textbf{44.10\%} \\
1  & 12 & 82 & 334,639 & 18.07\% & 28.32\% & 53.61\% \\
2  & 1  & 7  & 6,406   & 32.55\% & 41.06\% & 26.40\% \\
3  & 2  & 13 & 319,854 & 24.44\% & 34.63\% & 40.94\% \\
4  & 2  & 21 & 13,486  & 21.55\% & 51.16\% & 27.28\% \\
5  & 1  & 3  & 1,428   & 28.78\% & 23.95\% & 47.27\% \\
6  & 2  & 14 & 27,085  & 16.72\% & 62.70\% & 20.58\% \\
7  & 1  & 6  & 3,662   & 36.07\% & 21.44\% & 42.49\% \\
8  & 2  & 18 & 42,853  & 24.17\% & 42.14\% & 33.69\% \\
9  & 1  & 7  & 4,353   & 8.73\%  & 12.01\% & 79.26\% \\
10 & 4  & 68 & 41,797  & 10.55\% & 56.50\% & 32.94\% \\
11 & 1  & 4  & 526     & 15.97\% & 54.37\% & 29.66\% \\
12 & 1  & 17 & 17,489  & 13.48\% & 59.91\% & 26.61\% \\
\bottomrule
\end{tabular}
\end{table*}

The clustering produced one cluster with substantially broader coverage than the others.
Cluster~0 contains 48 of the 78 functional categories, corresponding to 61.54\% of all categories.
These categories contain 472 of the 732 project instances, corresponding to 64.48\% of all projects.
They also contain 743,407 of the 1,556,985 classified Java files, corresponding to 47.75\% of all files.

The three EBC proportions form one composition and must be interpreted together.
A ternary plot represents each functional category as one point whose position is determined by its Entity, Boundary, and Control proportions.
A point closer to a vertex has a higher proportion of the corresponding role.
Figure~\ref{fig:S4-cluster-ternary} highlights the dominant cluster and separates it from the non-dominant categories.
This presentation shows the location of the recurring composition without requiring the individual non-dominant clusters to be distinguished visually.

The ternary plot shows both concentration and variation in the category-level compositions.
The dominant-cluster categories occupy a concentrated region around the marked mean.
The non-dominant categories are more widely distributed, although several remain close to the dominant cluster.
Most categories occur in the lower part of the plot because their Entity proportions are smaller than their Boundary and Control proportions.
The figure therefore provides visual support for a recurring composition while showing that the pattern does not cover every functional category.

The dominant cluster has a mean composition of 13.25\% Entity, 42.65\% Boundary, and 44.10\% Control.
The median composition is 12.85\% Entity, 43.30\% Boundary, and 43.18\% Control.
The similarity between the mean and median indicates that both measures identify the same central composition.

The proportions within Cluster~0 remain concentrated around its central composition.
Table~\ref{tab:S4-recurring-composition} reports the distribution of each EBC role among the 48 functional categories.
The interquartile ranges are 5.24 percentage points for Entity, 7.87 percentage points for Boundary, and 10.42 percentage points for Control.

\begin{table}[htbp]
\centering
\caption{Distribution of classifier-derived EBC proportions within Cluster~0.}
\label{tab:S4-recurring-composition}
\begin{tabular}{lccc}
\toprule
\textbf{Statistic} &
\textbf{Entity} &
\textbf{Boundary} &
\textbf{Control} \\
\midrule
Mean               & 13.25\% & 42.65\% & 44.10\% \\
Standard deviation & 3.64\%  & 5.25\%  & 6.95\% \\
\midrule
Minimum            & 9.77\%  & 36.50\% & 37.18\% \\
First quartile     & 10.58\% & 38.55\% & 38.34\% \\
Median             & 12.85\% & 43.30\% & 43.18\% \\
Third quartile     & 15.81\% & 46.42\% & 48.76\% \\
Maximum            & 21.50\% & 51.77\% & 63.03\% \\
\midrule
Interquartile range & 5.24\% & 7.87\%  & 10.42\% \\
\bottomrule
\end{tabular}
\end{table}

The central composition remains similar under nearby DBSCAN parameter values.
For \(\varepsilon=0.5\), the largest cluster has a mean composition of 12.02\% Entity, 42.39\% Boundary, and 45.59\% Control.
For \(\varepsilon=0.7\), the corresponding composition is 14.01\%, 41.04\%, and 44.95\%.
The mean of each role therefore changes by no more than 1.99 percentage points across \(\varepsilon\) values of 0.5, 0.6, and 0.7.
This result supports the stability of the central composition, but it does not imply identical cluster membership across the three settings.

The dominant cluster represents the recurring composition identified in this study.
This composition is termed the \emph{EBC Proportionality Law}.
The term \emph{Law} denotes an observed empirical regularity within the analysed Java corpus rather than a universal or required software architecture.

The EBC Proportionality Law does not represent every functional category.
The remaining 30 categories are distributed across 12 smaller clusters and contain 260 project instances and 813,578 classified Java files.
The dominant cluster therefore contains most functional categories and project instances, but slightly fewer than half of the classified files.
These remaining compositions are examined as variations in RQ3.

\begin{center}
\fbox{
\parbox{0.94\linewidth}{
\textbf{Answer to RQ2.}
Under the chosen DBSCAN parameters, the largest cluster contains 48 of the 78 functional categories and 472 of the 732 project instances.
Its mean composition is 13.25\% Entity, 42.65\% Boundary, and 44.10\% Control.
The Law is broadly represented across categories and projects, but it does not cover every category or most classified files.
}}
\end{center}

\subsection{RQ3: Variations Across Functional Categories}
\label{S4:RQ3}

RQ3 examines the functional-category variations from the EBC Proportionality Law.
The 30 categories outside the dominant cluster contain 260 project instances and 813,578 classified Java files.
These categories form 12 non-dominant clusters.
The analysis first classifies their quantitative differences and then examines selected architectural interpretations.

\subsubsection{Classification of Variations}

The variations were classified relative to the distribution of Cluster~0.
Table~\ref{tab:S4-variation-thresholds} reports the thresholds used for each EBC role.
A value between the first and third quartiles was classified as \textsc{Typical}.
Values below this interval were classified as \textsc{Low} or \textsc{Very Low}, while values above it were classified as \textsc{High} or \textsc{Very High}.
These classifications describe differences from Cluster~0 and do not represent architectural quality.

\begin{table}[htbp]
\centering
\caption{Thresholds used to classify variations from the EBC Proportionality Law.}
\label{tab:S4-variation-thresholds}
\begin{tabular}{lccc}
\toprule
\textbf{Threshold} &
\textbf{Entity} &
\textbf{Boundary} &
\textbf{Control} \\
\midrule
Lower IQR bound & 2.72\%  & 26.75\% & 22.72\% \\
First quartile  & 10.58\% & 38.55\% & 38.34\% \\
Third quartile  & 15.81\% & 46.42\% & 48.76\% \\
Upper IQR bound & 23.66\% & 58.22\% & 64.38\% \\
\bottomrule
\end{tabular}
\end{table}

The non-dominant clusters vary in several directions.
Their exact mean proportions are reported in Table~\ref{tab:S4-cluster-summary}.
Table~\ref{tab:S4-relative-patterns} applies the IQR classifications to those means.
The symbolic scale provides a compact comparison of the non-dominant clusters.
Table~\ref{tab:S4-relative-patterns} uses \ebcVeryLow{} for Very Low, \ebcLow{} for Low, \ebcTypical{} for Typical, \ebcHigh{} for High, and \ebcVeryHigh{} for Very High.
Orange symbols represent proportions below the recurring range, grey represents the recurring range, and blue represents proportions above the recurring range.
The symbols describe direction and magnitude relative to the dominant classifier-derived EBC composition and do not indicate architectural quality.
This presentation separates the numerical clustering results from their interpretation.

\begin{table}[htbp]
\centering
\caption{Symbolic EBC patterns of the non-dominant clusters.}
\label{tab:S4-relative-patterns}
\small
\begin{tabular}{ccccc}
\toprule
\textbf{Cluster} &
\textbf{Categories} &
\textbf{Entity} &
\textbf{Boundary} &
\textbf{Control} \\
\midrule
1  & 12 & \ebcHigh     & \ebcLow      & \ebcHigh     \\
\midrule
2  & 1  & \ebcVeryHigh & \ebcTypical  & \ebcLow      \\
3  & 2  & \ebcVeryHigh & \ebcLow      & \ebcTypical  \\
4  & 2  & \ebcHigh     & \ebcHigh     & \ebcLow      \\
5  & 1  & \ebcVeryHigh & \ebcVeryLow  & \ebcTypical  \\
6  & 2  & \ebcHigh     & \ebcVeryHigh & \ebcVeryLow  \\
7  & 1  & \ebcVeryHigh & \ebcVeryLow  & \ebcTypical  \\
8  & 2  & \ebcVeryHigh & \ebcTypical  & \ebcLow      \\
9  & 1  & \ebcLow      & \ebcVeryLow  & \ebcVeryHigh \\
10 & 4  & \ebcLow      & \ebcHigh     & \ebcLow      \\
11 & 1  & \ebcHigh     & \ebcHigh     & \ebcLow      \\
12 & 1  & \ebcTypical  & \ebcVeryHigh & \ebcLow      \\
\bottomrule
\end{tabular}

\vspace{3pt}
\begin{minipage}{0.92\linewidth}
\footnotesize
\textit{Symbols:}
\ebcVeryLow{} Very Low;
\ebcLow{} Low;
\ebcTypical{} Typical;
\ebcHigh{} High;
\ebcVeryHigh{} Very High.
\end{minipage}
\end{table}

\subsubsection{Alternative Proportionality Pattern}

The largest non-dominant cluster represents an alternative proportionality pattern.
Cluster~1 contains 12 functional categories (15.38\%), 82 project instances (11.20\%), and 334,639 classified Java files (21.49\%).
Its mean composition is 18.07\% Entity, 28.32\% Boundary, and 53.61\% Control.
This composition combines high Entity, low Boundary, and high Control proportions.
Its coverage across 12 categories distinguishes it from the smaller non-dominant clusters.

\subsubsection{Specialised Proportionality Patterns}

The remaining non-dominant clusters represent specialised proportionality patterns.
Clusters~2--12 each contain between one and four functional categories.
Together, they contain 18 functional categories (23.08\%), 178 project instances (24.32\%), and 478,939 classified Java files (30.76\%).
Their combined coverage is distributed across 11 different patterns rather than one shared composition.

The specialised patterns differ in several EBC directions.
Clusters~2, 3, 5, 7, and 8 are primarily distinguished by very high Entity proportions.
Clusters~6 and 12 have very high Boundary proportions, while Cluster~9 has a very high Control proportion.
Clusters~4, 10, and 11 contain combined differences across multiple roles.
These results show that variation from Cluster~0 does not occur in one common direction.

\subsubsection{Architectural Consistency of Selected Variations}

The architectural assessment presents five illustrative categories in the main text.
These categories cover different directions of EBC variation and include both consistent and less consistent interpretations.
Table~\ref{tab:S4-architectural-interpretation} presents the selected cases.
The assessment of the remaining non-dominant functional categories, with supporting references and role-specific interpretations, is available in the replication package.

\begin{table*}[htbp]
\centering
\renewcommand{\arraystretch}{1.5} 
\caption{Selected architectural interpretations of variations from the EBC Proportionality Law.}
\label{tab:S4-architectural-interpretation}
\small
\begin{tabularx}{\textwidth}{
@{}
p{0.13\textwidth}
p{0.11\textwidth}
X
p{0.20\textwidth}
@{}}
\toprule
\textbf{Category} &
\textbf{Observed\newline pattern} &
\textbf{Architectural characteristic} &
\textbf{Interpretation} \\
\midrule

High\newline Performance &
E: \ebcLow{}\newline
B: \ebcVeryLow{}\newline
C: \ebcVeryHigh{} &
Runtime systems coordinate parallel scheduling, computation, communication, data placement, load balancing, and resource management~\cite{aumage2021task,sterling2017survey}. &
The Control \ebcVeryHigh{} proportion is consistent. \\

Bean Mapping &
E: \ebcVeryHigh{}\newline
B: \ebcTypical{}\newline
C: \ebcLow{} &
Object-relational mapping places entities and their corresponding objects at the centre of the representation model~\cite{o2008object}. &
The Entity \ebcVeryHigh{} proportion is consistent. \\

Game\newline Development &
E: \ebcHigh{}\newline
B: \ebcVeryHigh{}\newline
C: \ebcVeryLow{} &
Game engines provide rendering, input, sound, networking, physics, artificial intelligence, and event-management components~\cite{freiknecht2016game,vohera2021game,zhang2024auxiliary}. &
The Boundary \ebcVeryHigh{} proportion is consistent, but the Control \ebcVeryLow{} proportion is less consistent. \\

Code Generators &
E: \ebcHigh{}\newline
B: \ebcHigh{}\newline
C: \ebcLow{} &
Code-generation systems process external inputs, construct intermediate representations, and coordinate generation activities~\cite{chennamsetty2025bridging,zhao2025towards}. &
The Entity \ebcHigh{} and Boundary \ebcHigh{} proportions are consistent, but the Control \ebcLow{} proportion is less consistent. \\

Financial &
E: \ebcVeryHigh{}\newline
B: \ebcVeryLow{}\newline
C: \ebcTypical{} &
Financial systems interact with external trading venues, client activities, and multiple external data sources~\cite{munar2014big,tian2015latency}. &
The Boundary \ebcVeryLow{} proportion is less consistent. \\

\bottomrule
\end{tabularx}

\vspace{3pt}
\begin{minipage}{0.98\textwidth}
\footnotesize
\textit{Symbols:}
\ebcVeryLow{} Very Low;
\ebcLow{} Low;
\ebcTypical{} Typical;
\ebcHigh{} High;
\ebcVeryHigh{} Very High.
Orange indicates a proportion below the recurring range, grey indicates the recurring range, and blue indicates a proportion above the recurring range.
The symbols indicate relative position rather than architectural quality.
\end{minipage}
\end{table*}

The selected cases contain both consistent and less consistent results.
High Performance provides a Control-oriented example because runtime coordination is consistent with its very high Control proportion.
Bean Mapping provides an Entity-oriented example because object representations are consistent with its very high Entity proportion.
These cases show clear agreement between the observed variation and the architectural responsibilities described in the literature.

The combined variations produce more mixed interpretations.
Game Development contains interaction-oriented components that are consistent with its very high Boundary proportion, but its very low Control proportion is less consistent with the coordination performed by game engines.
Code Generators show a similar result because their high Entity and Boundary proportions are plausible, while their low Control proportion is less consistent with coordinated generation activities.
A functional category can therefore contain both consistent and less consistent EBC dimensions.

The selected evidence does not explain every observed variation.
The very low Boundary proportion of Financial software is difficult to align with its interaction with trading venues, clients, and external data sources.
Other non-dominant categories may also require evidence from their specific repositories rather than general descriptions of their functional category.
The architectural assessment therefore provides illustrative and role-specific interpretations rather than validation of all 30 categories.

\begin{center}
\fbox{
\parbox{0.94\linewidth}{
\textbf{Answer to RQ3.}
The 30 functional categories \textbf{outside} the dominant cluster form one alternative pattern and several specialised patterns.
Their variations include Entity-heavy, Boundary-heavy, Control-heavy, and combined compositions under the classifier.
The selected architectural evidence is consistent with some individual variations, while other variations have mixed or less consistent explanations.
Since the underlying labels have only limited agreement, the links between functional characteristics and EBC variation are partial and role-specific.
}}
\end{center}

\section{Discussion and Implications}
\label{S5:Discussion}

The results show one recurring pattern and several variations in classifier-derived EBC proportions.
RQ1 describes the distribution of classifier-derived EBC proportions, RQ2 identifies the recurring pattern, and RQ3 examines the variations.
This section explains what these findings mean for researchers, practitioners, and software engineering educators.
Each Java file received one representative EBC label, so the proportions describe the distribution of file labels rather than every responsibility within each file.
The EBC Proportionality Law describes an observed pattern in these classifier-derived measurement, not a required design or a measure of software quality.

\subsection{Implications for Researchers}
\label{S5:Researchers}

\subsubsection{RQ1: EBC Distribution as a Research Baseline}

The average classifier-derived EBC proportions are similar at the project and functional-category levels.
At both levels, Entity accounts for about 15\% of the classified files, while Boundary and Control account for much larger and similar shares.
The central balance is therefore visible under both ways of grouping the files.
Earlier studies examined architectural roles in methods, classes, or smaller collections of projects~\cite{dragan2010automatic,moreno2012jstereocode,hoquang2022role}.
The present results give researchers a wider, classifier-derived basis for comparison.

Functional categories show less variation in EBC proportions than individual projects.
Pooling files from several projects can reduce differences that are visible between those projects.
The smaller spread does not mean that all projects in one category have similar architectures.
A functional category describes a project's main purpose rather than its internal structure~\cite{forward2008taxonomy,nagappan2013diversity}.
Researchers should therefore report how projects are grouped when comparing EBC proportions.

The file-level measurement limits what the proportions mean.
Each file counts once, regardless of its length or the number of responsibilities it contains.
The expert comparison also showed low recall for Entity, so the measured Entity proportion may partly reflect Entity files missed by the classifier.
Future studies should compare file-level proportions with measurements based on lines of code, classes, and methods.
Such comparisons would show whether the central balance remains similar when the measurement unit changes.

\subsubsection{RQ2: Scope of the EBC Proportionality Law}

The EBC Proportionality Law recurs across most categories and project instances, but not across most classified files.
The dominant cluster contains 48 of the 78 functional categories and 472 of the 732 project instances.
These projects contain 47.75\% of the classified Java files.
The strength of the recurrence therefore depends on whether it is counted by categories, projects, or files.
Researchers should state which of these units supports a claim about the Law's coverage.

The Law describes a recurring balance of representative roles across files.
The mean proportions of the dominant cluster are 13.25\% Entity, 42.65\% Boundary, and 44.10\% Control.
All three roles contribute to the pattern, although they do not contribute equally.
These proportions do not measure how much of each responsibility occurs within an individual file or class.
The Law provides a point of comparison for research, not evidence that this balance is the best design.

The scope and effects of the Law remain open questions.
The pattern was identified in open-source Java projects and may differ in other languages or software collections.
Studying several releases could show whether projects remain near the Law or change over time; earlier work has examined changes in architectural-role distributions across versions~\cite{hoquang2022role}.
The present study also does not test whether distance from the Law is related to defects or maintenance effort.
Those relationships require separate studies that account for project differences and classifier uncertainty.

\subsubsection{RQ3: Meaning of EBC Variations}

The non-dominant clusters show that the recurring balance is not the only pattern in the dataset.
Thirty functional categories fall outside the dominant cluster.
Some share an alternative pattern, while others form smaller specialised patterns.
These variations do not contradict the Law because the Law describes a recurring pattern rather than a rule for every category.

Published descriptions of functional purpose explains some variations more clearly than others.
The high Control proportion in High Performance software is consistent with its coordination tasks.
The high Entity proportion in Bean Mapping software is consistent with its focus on object representation.
The selected results for Game Development and Code Generators have mixed explanations, while the low Boundary proportion in Financial software is harder to explain from its common functions.
A category's purpose therefore helps explain some roles but not the complete EBC pattern.

Variations need further evidence from the projects that produce them.
Projects in the same category may use different frameworks, designs, and ways of organising files.
Classifier errors may also affect their measured proportions.
The present results cannot show whether a variation lasts across releases or relates to defects.
For researchers, variation is a finding to explain, not evidence that a project has an incorrect design.

\subsection{Implications for Practitioners}
\label{S5:Practitioners}

\subsubsection{RQ1: EBC Proportions as a Project Summary}

Project-level classifier-derived EBC proportions give a short summary of representative roles across a project's Java files.
The three numbers show how often files were assigned to Entity, Boundary, and Control.
They provide an initial view of the project's role balance without requiring every file to be read.
They do not show the full architecture or every responsibility within an individual file.

EBC proportions can complement architectural documentation.
Architecture documentation explains the main structures and responsibilities of a system~\cite{bass2021software,hasselbring2018software}.
The proportions add a small summary that can be calculated again when the code changes.
The underlying file labels can then guide closer inspection of particular files.
The summary should include the number of classified files and a note about classifier limitations.

Repeated measurement can show changes in a project's role balance.
A change between releases means that the distribution of representative file labels has changed.
It does not, by itself, show that the architecture has improved or drifted.
Practitioners would need to examine the changed files and the purpose of the release before drawing that conclusion.

\subsubsection{RQ2: The EBC Proportionality Law as a Reference}

The Law provides a common pattern against which a project can be compared.
Its recurrence across most categories and project instances makes it broader than a pattern taken from one project alone.
A comparison shows which roles occur more or less often in the project than in the recurring pattern.
The Law is not a target percentage because a project's purpose may require a different balance.

A difference from the Law provides a reason to ask questions about the project.
Practitioners can compare the project with earlier releases and with other projects in the same functional category.
They can then inspect the files that contribute to the difference and consider whether the role balance fits the project's purpose.
This use supports architectural inspection rather than automatic judgment.

The Law does not predict software quality in the present study.
The results do not show that projects closer to the Law have fewer defects or need less maintenance.
Distance from the Law should therefore not be used on its own to set testing priorities or report a defect risk.
Such uses would require direct evidence linking EBC proportions with quality outcomes.

\subsubsection{RQ3: EBC Variations as Reasons for Inspection}

Some variations have a plausible connection with software function.
For example, High Performance software has a high Control proportion, while Bean Mapping software has a high Entity proportion.
Other selected variations are harder to explain.
A difference from the Law should therefore lead to examination of the project's purpose and design, not an immediate decision that the design is wrong.

An unexpected variation can guide a closer review.
Practitioners can inspect the files assigned to roles that occur unusually often or unusually rarely.
They should also check uncertain predictions because an unusual proportion may partly reflect classification errors.
Project-level proportions reveal a difference in role balance; the source files are needed to explain it.

An intentional variation does not need to be removed.
A project may require a role balance that differs from the recurring pattern.
If inspection shows that the difference fits the project's purpose, its reason can be recorded in the architectural documentation.
If the difference remains unexplained, further review is reasonable, but the variation alone is not evidence of a defect.

\subsection{Implications for Software Engineering Educators}
\label{S5:Educators}

\subsubsection{RQ1: Teaching EBC Roles}

The observed distribution gives students an example of EBC roles in real source code.
Entity is less frequent than Boundary and Control among the representative file labels in this corpus.
Entity concerns data and domain state, Boundary concerns interaction with users and external systems, and Control concerns workflow coordination~\cite{jacobson1993object,wirfsbrock2003object}.
Students can examine these roles in files and then see how their counts form project-level proportions.

A small project can turn the distribution into a teaching exercise.
Students can assign EBC roles to its files, count the labels, and compare the result with other projects.
Earlier work examined the use of automatic EBC labelling and explanation in software engineering education~\cite{fudholi2026enhancing}.
The present results add project-level and category-level comparisons to that type of exercise.

File-level labels also provide a way to discuss mixed responsibilities.
One file may contain several classes or methods with different roles, although the classifier gives it only one representative label.
Students can inspect such files and question whether the label captures their main responsibility.
Because the classifier showed limited agreement with expert labels, its predictions should be examined rather than treated as an answer key.

\subsubsection{RQ2: Teaching the EBC Proportionality Law}

The coverage of the Law shows students why the unit being counted matters.
The pattern covers most functional categories and project instances but fewer than half of the classified files.
Students can compare these three measures of coverage and explain why they support different statements about how common the Law is.
This exercise makes the meaning of a recurring pattern more precise.

The Law also shows the difference between an observation and a design rule.
Students can compare a project's proportions with the recurring pattern and look for reasons behind any difference.
They should consider how the files were labelled and whether the project's purpose calls for a different balance.
The Law is an empirical reference for this discussion, not a template that every project should follow.

Several releases of one project can provide a further exercise.
Students can calculate EBC proportions for each release and inspect changes in the code and release notes.
They can discuss possible reasons for a change without assuming that movement toward the Law means improvement.
The present study does not test that relationship, so the exercise should focus on explaining measured changes.

\subsubsection{RQ3: Teaching EBC Variations}

The non-dominant patterns show students that a difference from the Law is not automatically a design failure.
The results for High Performance and Bean Mapping have plausible links to their common tasks.
Other selected categories have patterns that are only partly explained by their functions.
Students can compare these cases and decide what evidence supports each explanation.

A functional-category name is only a starting point for understanding a variation.
Students can inspect projects within a category and consider their frameworks, files, and design choices.
They can then assess whether the category-level explanation fits the actual code.
This comparison teaches the difference between a project's general purpose and its particular implementation.

Architectural decisions need to be judged in their context.
Students can ask whether an unusual role balance follows from the project's requirements or from uncertain file labels.
A well-explained variation need not be changed merely to match the Law.
An unexplained variation calls for closer study rather than automatic correction.

\section{Threats to Validity}
\label{S6:Threats}
Several validity threats affect the interpretation of the findings.
These threats concern the representation of EBC roles, the analytical decisions, the coverage of the dataset, and the reproducibility of the study.
The following subsections describe the main threats, the measures taken to reduce them, and the limitations that remain.

\subsection{Construct Validity}
\label{S6:Construct}
Construct validity concerns how EBC roles and software functionality are represented.
EBC was originally defined for objects and classes, while this study assigns one representative role to each Java file.
A file may contain several classes, methods, and architectural responsibilities.
The assigned label should therefore be interpreted as the representative role detected for the file rather than a complete description of its contents.

Automated classification represents the main threat to the EBC measurements.
The classifier was trained using LLM-generated weak labels and showed limited agreement with the expert annotations.
This limitation may introduce classification errors into the project- and category-level proportions, particularly through the under-identification of Entity files.
The study therefore reports the expert comparison and treats the resulting proportions as classifier-derived measurements.

Functional categories provide an approximate representation of software purpose.
The categories describe the main purpose assigned to each project in the Awesome Java collection, but a project may implement several functions or contain components with different responsibilities.
The categories provide a consistent basis for comparison, although they do not provide a complete description of each project’s architecture.
The architectural interpretations in RQ3 should therefore be treated as illustrative rather than complete validation.

\subsection{Internal Validity}
\label{S6:Internal}

Internal validity concerns whether the study procedure introduced unintended differences into the results.
Repository availability and filtering decisions may affect the final corpus.
The exclusion criteria were applied before EBC annotation and did not depend on the proportionality of a project.
However, inaccessible repositories could not be examined and may differ from the retained projects.

Automated processing may introduce errors during file collection, classification, and aggregation.
The study recorded the number of files at each main processing stage and excluded 13 paths that did not represent readable Java files.
The same classification and aggregation procedure was applied to every retained project.
These controls reduce processing differences but cannot guarantee that every source file was handled without error.

The construction of the evaluation and project datasets creates additional limitations.
The expert labels were not used during classifier training, but the expert-annotated files came from a project represented in the weak-label corpus.
The expert comparison is therefore not fully independent at the project level.
In addition, two major versions of one parent project were treated as separate project instances, which gives that parent project two observations in the same functional category.

\subsection{Conclusion Validity}
\label{S6:Conclusion}

Conclusion validity concerns whether the findings support the conclusions drawn from the analysis.
The study is descriptive and does not establish causal relationships between EBC proportionality and software quality.
Entity, Boundary, and Control proportions also sum to 100\%, meaning that changes in the three roles are not independent.

The aggregation procedure may affect the category-level results.
Category proportions were calculated by pooling classified files from all projects in the category.
Large projects therefore contribute more strongly than small projects, and part of the lower category-level dispersion may result from aggregation.

The dominant composition depends on which clustering procedure has been selected; nearby DBSCAN parameter values produced similar central proportions, but they did not necessarily produce identical cluster memberships.
Cluster~0 covers most functional categories and projects but contains 47.75\% of the classified files.
The finding therefore supports a recurring empirical pattern rather than a universal composition.

The architectural assessment provides limited evidence about the meaning of the variations.
The selected examples show that some EBC differences are consistent with common functional responsibilities, while other differences receive mixed support.
These examples provide possible explanations but do not establish that functionality causes the observed variations.

\subsection{External Validity}
\label{S6:External}

External validity concerns whether the findings apply beyond the analysed corpus.
The dataset contains open-source Java projects collected from the Awesome Java collection.
The results may differ for other programming languages, proprietary systems, industrial applications, or projects that are not included in curated software collections.

The functional categories have different levels of representation.
Some categories contain many projects, while others contain only one or a few projects.
The classifier-derived EBC proportions of smaller categories may depend strongly on the architecture of individual projects.
The dominant EBC composition should therefore be interpreted within the category and project coverage reported in this study.

The dataset represents the projects at particular points in their development.
Software repositories and their EBC proportions may change as new functionality is added and existing code is restructured.
The findings describe the acquired project versions and do not establish that the same proportions remain stable over time.

\subsection{Reliability and Reproducibility}
\label{S6:Reliability}

Reliability concerns whether the study can be repeated using the same procedure.
Weak-label generation used the hosted \texttt{gpt-5-mini} model and the API defaults available during the annotation period.
Changes to the hosted model or its default settings may produce different labels in a later execution.
The model identifier, prompt, response schema, annotation dates, and raw responses were retained, although the exact client-library version was not recorded.

The analysis also depends on several software tools and parameter settings.
The feature extraction, classifier training, clustering, and statistical calculations must use the same implementations to reproduce the reported results.
The replication package therefore documents the analysis parameters and provides the processing scripts, trained classifier, predictions, and aggregated results.

Repository availability creates an additional reproducibility risk.
Open-source repositories may be changed, renamed, made private, or removed after data collection.
Repository locations, acquired versions, category assignments, and exclusion records should therefore accompany the replication materials.
Private API credentials are not required to inspect the stored weak labels and are excluded from the package.

\section{Conclusion}
\label{S7:Conclusion}

This study examined how Entity, Boundary, and Control responsibilities are distributed across Java software.
The analysis covered 732 project instances, 78 functional categories, and 1,556,985 classified Java files.
File-level EBC predictions were aggregated to describe proportionality at the project and functional-category levels.
This design allowed EBC roles to be examined as a software composition rather than only as labels assigned to individual source-code elements.

The results identify both recurrence and variation in classifier-derived EBC proportionality.
Entity forms the smallest proportion at the project and functional-category levels, while Boundary and Control form larger and similar proportions.
The dominant cluster contains 48 functional categories and 472 project instances, with a mean composition of 13.25\% Entity, 42.65\% Boundary, and 44.10\% Control.
This recurring composition is defined as the EBC Proportionality Law.
The remaining categories form one alternative pattern and several specialised patterns with different Entity, Boundary, and Control compositions.

The EBC Proportionality Law provides a new empirical perspective on architectural responsibilities.
It provides researchers with a proposition for further testing, practitioners with a reference for architectural inspection, and educators with a connection between EBC concepts and real software systems.
The Law does not require every software system to follow the same proportions.
It provides an empirical comparison point from which functionally necessary and architecturally meaningful variations can be examined.

The findings remain limited by the use of automated file-level classification.
The classifier showed limited agreement with expert annotations, and one file may contain several classes or architectural responsibilities.
Future research should examine the EBC Proportionality Law using independent datasets, other programming languages, stronger expert validation, and class- or method-level analysis.
Longitudinal studies can also examine how EBC proportionality changes during software evolution and whether these changes are related to bug-fixing activity.

The mixed-role annotations suggest that file-level analysis may combine responsibilities implemented by different classes or methods.
Future research can examine EBC roles at class and method levels and determine how these finer-grained roles combine into file-level and project-level proportionality.
Such analysis may distinguish files with one representative responsibility from files that intentionally combine several architectural responsibilities.

\begin{acks}
This work was conducted as part of the first author's doctoral research, which is funded by the Indonesia Endowment Fund for Education (Lembaga Pengelola Dana Pendidikan, LPDP), Ministry of Finance of the Republic of Indonesia.

\end{acks}
\bibliographystyle{ACM-Reference-Format}
\bibliography{main}

@book{hofmeister2000applied,
title={Applied software architecture},
author={Hofmeister, Christine and Nord, Robert and Soni, Dilip},
year={2000},
publisher={Addison-Wesley Professional}
}

@inproceedings{medvidovic2010software,
title={Software architecture: foundations, theory, and practice},
author={Medvidovic, Nenad and Taylor, Richard N},
booktitle={Proceedings of the 32nd ACM/IEEE International Conference on Software Engineering-Volume 2},
pages={471--472},
year={2010}
}

@book{jacobson1993object,
  title={Object-oriented software engineering: a use case driven approach},
  author={Jacobson, Ivar},
  year={1993},
  publisher={Pearson Education India}
}

@book{larman2012applying,
  title={Applying UML and patterns: an introduction to object oriented analysis and design and interative development},
  author={Larman, Craig},
  year={2012},
  publisher={Pearson Education India}
}

@article{sas2023gitranking,
  title={GitRanking: A ranking of GitHub topics for software classification using active sampling},
  author={Sas, Cezar and Capiluppi, Andrea and Di Sipio, Claudio and Di Rocco, Juri and Di Ruscio, Davide},
  journal={Software: Practice and Experience},
  volume={53},
  number={10},
  pages={1982--2006},
  year={2023},
  publisher={Wiley Online Library}
}

@misc{awesomejava,
  author = {Akullpp},
  title = {Awesome Java},
  howpublished = {\url{https://github.com/akullpp/awesome-java}},
  note = {Accessed: April 2026}
}

@incollection{hasselbring2018software,
  title={Software architecture: Past, present, future},
  author={Hasselbring, Wilhelm},
  booktitle={The essence of software engineering},
  pages={169--184},
  year={2018},
  publisher={Springer}
}

@inproceedings{sas2021using,
  title={Using Structural and Semantic Information to Identify Software Components},
  author={Sas, Cezar and Capiluppi, Andrea},
  booktitle={2021 IEEE International Conference on Software Analysis, Evolution and Reengineering (SANER)},
  pages={546--550},
  year={2021},
  organization={IEEE}
}

@inproceedings{pangakis2024knowledge,
  title={Knowledge distillation in automated annotation: Supervised text classification with LLM-generated training labels},
  author={Pangakis, Nicholas and Wolken, Sam},
  booktitle={Proceedings of the Sixth Workshop on Natural Language Processing and Computational Social Science (NLP+ CSS 2024)},
  pages={113--131},
  year={2024}
}

@article{ke2017lightgbm,
  title={Lightgbm: A highly efficient gradient boosting decision tree},
  author={Ke, Guolin and Meng, Qi and Finley, Thomas and Wang, Taifeng and Chen, Wei and Ma, Weidong and Ye, Qiwei and Liu, Tie-Yan},
  journal={Advances in neural information processing systems},
  volume={30},
  year={2017}
}

@inproceedings{munar2014big,
  title={A big data financial information management architecture for global banking},
  author={Munar, Antoni and Chiner, Esteban and Sales, Ignacio},
  booktitle={2014 international conference on future internet of things and cloud},
  pages={385--388},
  year={2014},
  organization={IEEE}
}

@article{tian2015latency,
  title={Latency critical big data computing in finance},
  author={Tian, Xinhui and Han, Rui and Wang, Lei and Lu, Gang and Zhan, Jianfeng},
  journal={The Journal of Finance and Data Science},
  volume={1},
  number={1},
  pages={33--41},
  year={2015},
  publisher={Elsevier}
}

@inproceedings{o2008object,
  title={Object/relational mapping 2008: hibernate and the entity data model (edm)},
  author={O'Neil, Elizabeth J},
  booktitle={Proceedings of the 2008 ACM SIGMOD international conference on Management of data},
  pages={1351--1356},
  year={2008}
}

@misc{aumage2021task,
  title={Task-based performance portability in HPC},
  author={Aumage, Olivier and Carpenter, Paul and Benkner, Siegfried},
  year={2021}
}

@inproceedings{vohera2021game,
  title={Game engine architecture and comparative study of different game engines},
  author={Vohera, Chaitya and Chheda, Heet and Chouhan, Dhruveel and Desai, Ayush and Jain, Vijal},
  booktitle={2021 12th International Conference on Computing Communication and Networking Technologies (ICCCNT)},
  pages={1--6},
  year={2021},
  organization={IEEE}
}

@article{zhang2024auxiliary,
  title={An auxiliary development framework for lightweight RPG games based on Unity3D},
  author={Zhang, Bo and Shi, Huiping and Wang, Xinyu},
  journal={Computer Animation and Virtual Worlds},
  volume={35},
  number={1},
  pages={e2206},
  year={2024},
  publisher={Wiley Online Library}
}

@incollection{freiknecht2016game,
  title={Game Engines},
  author={Freiknecht, Jonas and Geiger, Christian and Drochtert, Daniel and Effelsberg, Wolfgang and D{\""o}rner, Ralf},
  booktitle={Serious Games: Foundations, Concepts and Practice},
  pages={127--159},
  year={2016},
  publisher={Springer}
}

@article{chennamsetty2025bridging,
  title={Bridging design and development: Building a generative AI platform for automated code generation},
  author={Chennamsetty, Chandra Shekar},
  journal={International Journal of Computer Technology and Electronics Communication},
  volume={8},
  number={2},
  pages={10420--10432},
  year={2025}
}

@article{zhao2025towards,
  title={Towards Realistic Project-Level Code Generation via Multi-Agent Collaboration and Semantic Architecture Modeling},
  author={Zhao, Qianhui and Zhang, Li and Liu, Fang and Cheng, Junhang and Wu, Chengru and Ai, Junchen and Meng, Qiaoyuanhe and Zhang, Lichen and Lian, Xiaoli and Song, Shubin and others},
  journal={ACM Transactions on Software Engineering and Methodology},
  year={2025},
  publisher={ACM New York, NY}
}

@article{sterling2017survey,
  title={A survey: runtime software systems for high performance computing},
  author={Sterling, Thomas and Anderson, Matthew and Brodowicz, Maciej},
  journal={Supercomputing Frontiers and Innovations},
  volume={4},
  number={1},
  pages={48--68},
  year={2017}
}

@book{wirfsbrock2003object,
  author    = {Rebecca Wirfs-Brock and Alan McKean},
  title     = {Object Design: Roles, Responsibilities, and Collaborations},
  publisher = {Addison-Wesley},
  year      = {2003},
  isbn      = {0-201-37943-0}
}

@book{jacobson1999unified,
  author    = {Ivar Jacobson and Grady Booch and James Rumbaugh},
  title     = {The Unified Software Development Process},
  publisher = {Addison-Wesley},
  address   = {Reading, MA},
  year      = {1999}
}

@techreport{reenskaug1979models,
  author      = {Trygve Reenskaug},
  title       = {Models--Views--Controllers},
  institution = {Xerox Palo Alto Research Center},
  year        = {1979},
  month       = dec,
  url         = {https://zenodo.org/records/3676092},
  note        = {Technical note, December 10, 1979}
}

@article{krasner1988cookbook,
  author  = {Glenn E. Krasner and Stephen T. Pope},
  title   = {A Cookbook for Using the Model--View--Controller
             User Interface Paradigm in Smalltalk-80},
  journal = {Journal of Object-Oriented Programming},
  volume  = {1},
  number  = {3},
  pages   = {26--49},
  year    = {1988},
  month   = {August/September},
  doi     = {10.5555/50757.50759},
  url     = {https://ics.uci.edu/~redmiles/ics227-SQ04/papers/KrasnerPope88.pdf}
}

@inproceedings{dragan2010automatic,
  author    = {Natalia Dragan and Michael L. Collard and Jonathan I. Maletic},
  title     = {Automatic Identification of Class Stereotypes},
  booktitle = {Proceedings of the 26th IEEE International Conference on Software Maintenance},
  pages     = {1--10},
  year      = {2010},
  doi       = {10.1109/ICSM.2010.5609703}
}

@inproceedings{moreno2012jstereocode,
  author    = {Laura Moreno and Andrian Marcus},
  title     = {{JStereoCode}: Automatically Identifying Method and Class Stereotypes in Java Code},
  booktitle = {Proceedings of the 27th IEEE/ACM International Conference on Automated Software Engineering},
  pages     = {358--361},
  year      = {2012},
  doi       = {10.1145/2351676.2351747}
}

@article{hoquang2022role,
  author  = {Truong Ho-Quang and Arif Nurwidyantoro and Satrio Adi Rukmono and Michel R. V. Chaudron and Fabian Fr{\"o}ding and Duy Nguyen Ngoc},
  title   = {Role Stereotypes in Software Designs and Their Evolution},
  journal = {Journal of Systems and Software},
  volume  = {189},
  pages   = {111296},
  year    = {2022},
  doi     = {10.1016/j.jss.2022.111296}
}

@inproceedings{forward2008taxonomy,
  author    = {Andrew Forward and Timothy C. Lethbridge},
  title     = {A Taxonomy of Software Types to Facilitate Search and Evidence-Based Software Engineering},
  booktitle = {Proceedings of the 2008 Conference of the Center for Advanced Studies on Collaborative Research},
  pages     = {171--191},
  year      = {2008},
  publisher = {Association for Computing Machinery},
  doi       = {10.1145/1463788.1463807}
}

@book{bass2021software,
  author    = {Len Bass and Paul Clements and Rick Kazman},
  title     = {Software Architecture in Practice},
  edition   = {4},
  publisher = {Addison-Wesley Professional},
  year      = {2021},
  isbn      = {978-0-13-688609-9}
}

@article{kawaguchi2006mudablue,
  author  = {Shinji Kawaguchi and Pankaj K. Garg and Makoto Matsushita and Katsuro Inoue},
  title   = {{MUDABlue}: An Automatic Categorization System for Open Source Repositories},
  journal = {Journal of Systems and Software},
  volume  = {79},
  number  = {7},
  pages   = {939--953},
  year    = {2006},
  month   = jul,
  doi     = {10.1016/j.jss.2005.06.044}
}

@inproceedings{nagappan2013diversity,
  author    = {Meiyappan Nagappan and Thomas Zimmermann and Christian Bird},
  title     = {Diversity in Software Engineering Research},
  booktitle = {Proceedings of the 2013 9th Joint Meeting on Foundations of Software Engineering},
  pages     = {466--476},
  year      = {2013},
  publisher = {Association for Computing Machinery},
  doi       = {10.1145/2491411.2491415}
}

@article{lehman1980programs,
  author  = {Meir M. Lehman},
  title   = {Programs, Life Cycles, and Laws of Software Evolution},
  journal = {Proceedings of the IEEE},
  volume  = {68},
  number  = {9},
  pages   = {1060--1076},
  year    = {1980},
  month   = sep,
  doi     = {10.1109/PROC.1980.11805}
}

@inproceedings{wheeldon2003power,
  author    = {Richard Wheeldon and Steve Counsell},
  title     = {Power Law Distributions in Class Relationships},
  booktitle = {Proceedings of the Third IEEE International Workshop on Source Code Analysis and Manipulation},
  pages     = {45--54},
  year      = {2003},
  publisher = {IEEE Computer Society},
  doi       = {10.1109/SCAM.2003.1238030}
}

@article{ratner2017snorkel,
  author  = {Alexander Ratner and Stephen H. Bach and
             Henry Ehrenberg and Jason Fries and
             Sen Wu and Christopher R{\'e}},
  title   = {Snorkel: Rapid Training Data Creation with Weak Supervision},
  journal = {Proceedings of the VLDB Endowment},
  volume  = {11},
  number  = {3},
  pages   = {269--282},
  year    = {2017},
  doi     = {10.14778/3157794.3157797}
}

@inproceedings{sas2023weaklabelling,
  author    = {Sas, Cezar and Capiluppi, Andrea},
  title     = {Weak Labelling for File-Level Source Code Classification},
  booktitle = {Proceedings of the 30th IEEE International Conference
               on Software Analysis, Evolution and Reengineering},
  pages     = {698--702},
  year      = {2023},
  publisher = {IEEE},
  doi       = {10.1109/SANER56733.2023.00074}
}

@misc{openai2026gpt5mini,
  author       = {{OpenAI}},
  title        = {{GPT-5 Mini} Model},
  year         = {2026},
  howpublished = {\url{https://developers.openai.com/api/docs/models/gpt-5-mini}},
  note         = {Accessed: 2026-09-02}
}

@article{salton1988term,
  author  = {Salton, Gerard and Buckley, Christopher},
  title   = {Term-Weighting Approaches in Automatic Text Retrieval},
  journal = {Information Processing \& Management},
  volume  = {24},
  number  = {5},
  pages   = {513--523},
  year    = {1988},
  doi     = {10.1016/0306-4573(88)90021-0}
}

@article{pedregosa2011scikit,
  author  = {Pedregosa, Fabian and Varoquaux, Ga{\"e}l and
             Gramfort, Alexandre and Michel, Vincent and
             Thirion, Bertrand and Grisel, Olivier and
             Blondel, Mathieu and Prettenhofer, Peter and
             Weiss, Ron and Dubourg, Vincent and
             VanderPlas, Jake and Passos, Alexandre and
             Cournapeau, David and Brucher, Matthieu and
             Perrot, Matthieu and Duchesnay, {\'E}douard},
  title   = {Scikit-Learn: Machine Learning in Python},
  journal = {Journal of Machine Learning Research},
  volume  = {12},
  number  = {85},
  pages   = {2825--2830},
  year    = {2011}
}

@inproceedings{fudholi2026enhancing,
  author    = {Fudholi, Dzikri Rahadian and Capiluppi, Andrea},
  title     = {Enhancing Software Engineering Education Through
               Explainable Automatic {EBC} Labelling},
  booktitle = {2026 IEEE International Conference on Software Analysis,
               Evolution and Reengineering -- Companion (SANER-C)},
  pages     = {106--110},
  year      = {2026},
  publisher = {IEEE},
  doi       = {10.1109/SANER-C67878.2026.00019}
}

@article{sokolova2009systematic,
  author  = {Sokolova, Marina and Lapalme, Guy},
  title   = {A Systematic Analysis of Performance Measures for
             Classification Tasks},
  journal = {Information Processing \& Management},
  volume  = {45},
  number  = {4},
  pages   = {427--437},
  year    = {2009},
  doi     = {10.1016/j.ipm.2009.03.002}
}

@book{aitchison1986statistical,
  author    = {Aitchison, John},
  title     = {The Statistical Analysis of Compositional Data},
  publisher = {Chapman and Hall},
  address   = {London},
  year      = {1986},
  isbn      = {9780412280603},
  doi       = {10.1007/978-94-009-4109-0}
}

@book{tukey1977exploratory,
  author    = {Tukey, John W.},
  title     = {Exploratory Data Analysis},
  publisher = {Addison-Wesley},
  address   = {Reading, Massachusetts},
  year      = {1977}
}

@inproceedings{ester1996density,
  author    = {Ester, Martin and Kriegel, Hans-Peter and Sander, J{\"o}rg and Xu, Xiaowei},
  title     = {A Density-Based Algorithm for Discovering Clusters in Large Spatial Databases with Noise},
  booktitle = {Proceedings of the Second International Conference on Knowledge Discovery and Data Mining},
  pages     = {226--231},
  year      = {1996},
  publisher = {AAAI Press}
}

@article{rousseeuw1987silhouettes,
  author  = {Rousseeuw, Peter J.},
  title   = {Silhouettes: A Graphical Aid to the Interpretation and Validation of Cluster Analysis},
  journal = {Journal of Computational and Applied Mathematics},
  volume  = {20},
  pages   = {53--65},
  year    = {1987},
  doi     = {10.1016/0377-0427(87)90125-7}
}










\end{document}